# The Hydrogen Atom: a Review on the Birth of Modern Quantum Mechanics

Luca Nanni

January 2015

## Abstract

The purpose of this work is to retrace the steps that were made by scientists of XXcentury, like Bohr, Schrodinger, Heisenberg, Pauli, Dirac, for the formulation of what today represents the modern quantum mechanics and that, within two decades, put in question the classical physics. In this context, the study of the electronic structure of hydrogen atom has been the main starting point for the formulation of the theory and, till now, remains the only real case for which the quantum equation of motion can be solved exactly. The results obtained by each theory will be discussed critically, highlighting limits and potentials that allowed the further development of the quantum theory.

## 1. Introduction

In scientific literature the discovery of hydrogen in atomic form is usually attributed to H. Cavendish and dates back to 1766 [1]. Since its discovery it was mainly characterized for its physico-chemical properties in order to study in detail its behavior in combustion reactions. It is only in 1855, after which Anders Angstrom published the results of his spectroscopic investigations on the line spectrum of hydrogen performed in 1852, that hydrogen atom became one of the most important research issues for the physicists of the time [2]. The line structure of its spectrum was not interpretable for the physics of XX century; on the other hand, just thinking that the electron was discovered by Thomson only in 1897 [3] to understand how the physicists of the time had not any opportunity to formulate an atomic theory able of explaining the line spectrum! However, this lack of knowledge prompted the physicists to acquire additional spectroscopic data and to improve the measuring apparatus in order to get information that, otherwise, would not be provided by the theory. It was just the way it was discovered the fine structure of hydrogen spectrum represented by the splitting of the spectral lines in presence of an external magnetic field (Zeeman effect, 1896).

The first measured performed by Angstrom pointed out that the spectrum was composed of three lines in the VIS band, a red-line at 6562.852 Å, a blue-green-line at 4861.33 Å and a violet line at 4340.47 Å. Subsequently, Angstrom improved the spectral resolution of its instrument finding that the violet line was formed by two distinct lines very close. In figure 1 is shown the original table published by Angstrom on the study of sun light spectrum:

| Raies | Sixième spectre | | Cinquième spectre | | Quatrième spectre | | Valeur moyenne de Longueur d'onde | Différence |
|---|---|---|---|---|---|---|---|---|
| | $m_6$ | $\lambda$ | $m_5$ | $\lambda$ | $m_4$ | $\lambda$ | | |
| | — | — | 918,0 | 4016,53 | 708,0 | 4016,94 | 4016,73 | 20 |
| | — | — | 1047,0 | 4004,62 | 810,0 | 04,80 | 04,71 | 9 |
| | — | — | — | — | 839,0 | 4001,36 | 4001,36 | — |
| | — | — | — | — | 869,0 | 3997,78 | 3997,78 | — |
| $\mathbf{H_1}$ | — | — | 1446,0 | 3967,76 | 1119,0 | 3968,00 | 67,88 | 12 |
| | — | — | *[1778,0 | 37,04] | — | — | — | — |
| $\mathbf{H_{11}}$ | — | — | 1823,0 | 3932,82 | — | — | 3932,82 | — |

**Figure 1**

In the years that followed the physicists began to notice some regularities of the spectrum and that some lines were related to other by an empirical equation. The first to undertake this study was Balmer who in 1885 proposed his empirical formula:

$$\lambda = B \left( \frac{m^2}{m^2 - 2^2} \right) \qquad (\mathbf{1.a})$$

where $\lambda$ is the wavelength of spectral line, B is a constant equal to 3645.6 Å (that corresponds to one of the lines lying in the UV band of the spectrum) and $m$ is an integer greater than 2 [4]. In 1888 the physicist J. Rydberg generalized the formula $(\mathbf{1.a})$ obtaining:

$$\frac{1}{\lambda} = \frac{4}{B} \left( \frac{1}{n'^2} - \frac{1}{n^2} \right) \quad n' = 1,2,3, \dots \ \ and \ n = 2,3,4, \dots \qquad (\mathbf{1.b})$$

To Balmer are dedicated the spectral lines of the VIS band of hydrogen spectrum and their position are listed in table 1:

| Balmer Series (n'=2) | |
|---|---|
| $n$ | $\lambda$ (nm) |
| 3 | 656.3 |
| 4 | 486.1 |
| 5 | 434.0 |
| 6 | 410.2 |
| 7 | 397.0 |

**Table 1**

To the physicist Lyman are dedicated the spectral lines lying in the UV band and were discovered in 1906-1914:

| Lyman Series (n'=1) | |
|---|---|
| *n* | λ (nm) |
| 2 | 122 |
| 3 | 103 |
| 4 | 97.3 |
| 5 | 95.0 |
| 6 | 93.8 |

**Table 2**

The lines in the IR band, instead, were discovered and studied by the German physicist F. Paschen in 1908; they partially overlap with those of the series attributed to Brackett (n'=4) and Pfund (n'=5), discovered in 1924 [5,6,7]. The positions of spectral lines for the three mentioned series are given in table 3:

| Paschen Series n'=3 | | Brackett Series n'=4 | | Pfund Series n'=5 | |
|---|---|---|---|---|---|
| *n* | λ (nm) | *n* | λ (nm) | *n* | λ (nm) |
| 4 | 1875 | --- | --- | --- | --- |
| 5 | 1282 | 5 | 4050 | --- | --- |
| 6 | 1094 | 6 | 2624 | 6 | 7460 |
| 7 | 1005 | 7 | 2165 | 7 | 4650 |
| 8 | 955 | 8 | 1944 | 8 | 3740 |
| 9 | 923 | 9 | 1817 | 9 | 3300 |
| 10 | 902 | --- | --- | 10 | 3040 |
| 11 | 887 | --- | --- | --- | --- |

**Table 3**

The hydrogen spectrum is completed by a final series due to the physicist C.J. Humphreys that in 1953 discovered lines in the microwave [8] band and whose positions are listed in table 4:

| Humphreys Series (n'=6) | |
|---|---|
| *n* | λ (nm) |
| 7 | 12400 |
| 8 | 7500 |
| 9 | 5910 |
| 10 | 5130 |
| 11 | 4670 |

**Table 4**

Nearly 60 years, in 1913, after its discovery the hydrogen spectrum finds its first theoretical explanation by N. Bohr that, using the Plank's concept of quanta, proposed the first quantum model in history, although it was still bound to the orthodoxy of classical physics and its concept of trajectory [9]. For completeness the structure of hydrogen spectra with the placement of the line series is shown in figure 2:

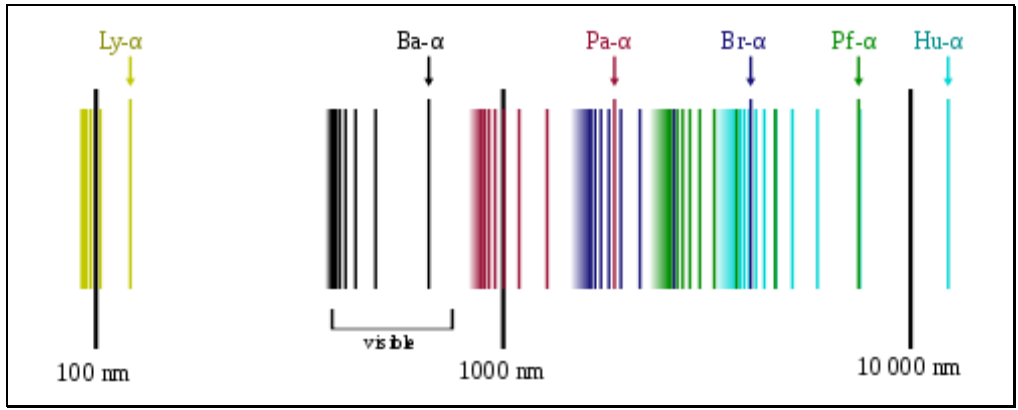

**Figure 2**

## 2. The Bohr Model

The model proposed by the physicists of the beginning '900 for the hydrogen atom is that planetary, where the electron of mass $m_e$ and charge $q_e$ rotates on a circular orbit about the atomic core of mass $m_N$ and charge $q_N$. The two particles interact by a central force of electrical nature. As a whole we have to study the motion of a particle in a conservative central field. According to the classical physics (electromagnetism) the rotation of the electron around the charged core leads to the progressively loss of energy as radiation till to the collapse of the atomic system! The experience, however, shows that the hydrogen atom is physically stable and the electron never fall on the nucleus.

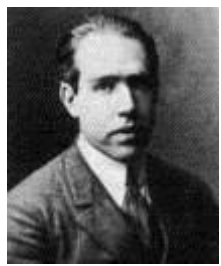

**Niels Bohr**

The planetary model of the hydrogen atom, according to the Lagrangian formalism, is characterized by six degrees of freedom. However, choosing the framework with the origin coinciding with the atomic center of mass, the degrees of freedom are reduced to three. The positions of the electron and core respect to the center of mass are given by the following vectors:

$$\boldsymbol{r}_e = \frac{m_e}{m_e + m_N} \boldsymbol{r}$$

$$\boldsymbol{r}_N = \frac{m_N}{m_e + m_N} \boldsymbol{r}$$

where $\boldsymbol{r}$ is the vector $(\boldsymbol{r}_e - \boldsymbol{r}_N)$ representing radius of the circular orbit. For such a system the kinetic energy is given by:

$$T = \frac{1}{2} m_e \dot{\boldsymbol{r}}_e{}^2 + \frac{1}{2} m_N \dot{\boldsymbol{r}}_N{}^2 = \frac{1}{2} \frac{m_e m_N}{m_e + m_N} \dot{\boldsymbol{r}}^2$$

where $(m_e m_N / m_e + m_N) = \mu$ is the reduced mass of the atom. The kinetic energy, then, can be rewritten as:

$$T = \frac{1}{2} \mu \dot{\boldsymbol{r}}^2$$

so that the Hamiltonian of the atom is the following:

$$H = T + V(\boldsymbol{r}) = \frac{1}{2} \mu \dot{\boldsymbol{r}}^2 + V(\boldsymbol{r})$$

H represents the total energy of the system under investigation. Since the Coulomb field in this model is conservative the total energy is a constant of motion. The Coulomb force between nucleus and electron is given by:

$$\boldsymbol{F} = \frac{1}{4\pi\varepsilon_0} q_e q_N \frac{\boldsymbol{r}}{|\boldsymbol{r}|^3} = -\frac{1}{4\pi\varepsilon_0} \frac{q_e^2}{r^2} \frac{\boldsymbol{r}}{|\boldsymbol{r}|}$$

Being the system stable the interaction force between electron and nucleus is balanced by the centrifugal one due to the circular motion:

$$\boldsymbol{F}_c = -m_e \omega^2 r \frac{\boldsymbol{r}}{|\boldsymbol{r}|}$$

Equalizing the two forces we get:

$$\frac{1}{4\pi\varepsilon_0} \frac{q_e^2}{r^2} = m_e \omega^2 r \qquad\qquad (\boldsymbol{2..a})$$

For a planetary model the angular momentum is conserved:

$$\boldsymbol{l} = \boldsymbol{r} \times \boldsymbol{p} = m_e \boldsymbol{v} \times \boldsymbol{r} = m_e \omega r^2 \boldsymbol{n} \qquad\qquad (\boldsymbol{2.b})$$

where $\boldsymbol{n}$ is the versor of the angular momentum (calculated according the rule of the right hand). The genial intuition of Bohr, although completely arbitrary, was that to quantize the modulus of the angular momentum according the following rule:

$$l = m_e v r = n \frac{h}{2\pi} = n \hbar \quad n \epsilon N \qquad\qquad (\boldsymbol{2.c})$$

from which it's obtained:

$$2\pi r = \frac{nh}{m_e v} = \frac{nh}{p} \qquad (2.d)$$

The circumference of the orbit $2\pi r$ is so divided in an integer number of $h/p$, that has the dimension of a length. This suggested to Bhor to associate to the orbit of the electron a standing wave with wave length $\lambda = (h/p)$. Bohr, even if unconsciously, arrived to the revolutionary idea of electron thought as material wave, anticipating of a decade the hypothesis of De Broglie [9,10]. Using the Bohr quantization rule we can calculate the energy of the hydrogen electron proceeding as follows.

The total energy of the planetary model of the atom is:

$$E = \frac{1}{2}\mu\omega^2 r^2 - \frac{1}{4\pi\varepsilon_0}\frac{q^2}{r}$$

From the **(2.b)** it's calculated the angular velocity:

$$\omega = \frac{l}{m_e r^2}$$

that substituted in the equation of the total energy gives:

$$E = \frac{1}{2}\mu\frac{l^2}{m_e^2 r^2} - \frac{1}{4\pi\varepsilon_0}\frac{q_e^2}{r} \qquad (2.e)$$

To calculate the position we use the **(2.a)**:

$$\frac{1}{4\pi\varepsilon_0}\frac{q_e^2}{r^2} = m_e\omega^2 r \quad \Rightarrow \quad \frac{1}{4\pi\varepsilon_0}\frac{q_e^2}{r^2} = m_e\omega^2 r = m_e\frac{l^2}{m_e^2 r^4}r$$

from which we get:

$$r = \frac{4\pi\varepsilon_0 l^2}{m_e q_e^2}$$

Substituting $r$ in the **(2.e)**:

$$E = \frac{1}{2}\mu\frac{l^2}{m_e^2}\frac{m_e^2 q_e^4}{(4\pi\varepsilon_0)^2 l^4} - \frac{1}{4\pi\varepsilon_0}\frac{q_e^2 m_e q_e^2}{4\pi\varepsilon_0 l^2} = \frac{1}{2}\mu\frac{q_e^4}{(4\pi\varepsilon_0)^2 l^2} - \frac{1}{(4\pi\varepsilon_0)^2}\frac{m_e q_e^4}{l^2}$$

Since $m_N \gg m_e$ it's possible to introduce the following approximation:

$$\mu = \frac{m_e m_N}{m_e + m_N} \cong \frac{m_e m_N}{m_N} = m_e$$

so that the center of mass of the system coincide with the atomic core. Using this result the energy above written becomes:

$$E = \frac{1}{2}m_e\frac{q_e^4}{(4\pi\varepsilon_0)^2 l^2} - \frac{1}{(4\pi\varepsilon_0)^2}\frac{m_e q_e^4}{l^2} = -\frac{1}{2(4\pi\varepsilon_0)^2}\frac{m_e q_e^4}{l^2}$$

Finally, substituting to $l$ the expression **(2.c)** we arrive at the following equation:

$$E_n == -\frac{1}{2(4\pi\varepsilon_0)^2}\frac{m_e q_e^4}{n^2 \hbar^2} \qquad (2.f)$$

The **(2..f)** shows that the electron can assume only quantized energy values depending on the integer number $n$. Increasing $n$, said principal quantum number, the spacing between the energy levels gets little and little. The radius of the orbits can be easily calculated inserting the quantized value of angular momentum in the expression of $r$ previously obtained:

$$r_n = \frac{4\pi\varepsilon_0}{m_e q_e^2} n^2 \hbar^2 \qquad (2.g)$$

The electron, like a standing wave, can rotate only on a circular orbit with quantized radius. Increasing $n$ the spacing of the orbit gets large and large. The equations obtained by Bohr were able to describe with high precision the hydrogen spectrum: each spectral line is due to an *electronic transition* from a given orbit to another one. In agreement with the Plank law $E = h\nu$, using the **(2.f)** we get just the Rydberg equation **(1.b)**:

$$\Delta E = E_{n'} - E_n = h\frac{c}{\lambda} \quad \rightarrow \quad \frac{1}{\lambda} = \frac{4}{B}\left(\frac{1}{n'^2} - \frac{1}{n^2}\right)$$

The way toward the quantum mechanics was definitely opened! The calculated wavelength values vs the experimental ones are listed in table 5:

| Spectral Line | Experimental Value | Theoretical Value |
|---|---|---|
| ----- | (nm) | (nm) |
| λ(n'=2, n=1) | 121.5 | 122.0 |
| λ(n'=3, n=1) | 102.5 | 103.0 |
| λ(n'=4, n=1) | 97.2 | 97.3 |
| λ(n'=2, n=3) | 656.1 | 656.3 |
| λ(n'=2, n=4) | 486.0 | 486.1 |
| λ(n'=3, n=4) | 1874.6 | 1875.0 |

**Table 5**

Starting from the **(2.b)** and the **(2.g)** is possible to obtain also the modulus of the electron velocity:

$$v = \frac{q_e^2}{4\pi\varepsilon_0 n\hbar}$$

which shows that increasing the principal quantum number the electron velocity decreases. If the electron velocity is relativistic then is reasonable thinking that increasing $n$ the mass tends to the rest one.

The Bohr model, however, is inadequate to explain both the fine structure of the hydrogen spectrum and the Zeeman effect. Moreover, it fails when is used to explain the spectrum of atoms with more than one electron. To be a new theoretical model it had a very small field of applicability; its importance, however, must be found in being able to introduce new physical concepts that were in strong contrast with the classical theory of motion!

## 3. Sommerfeld Conditions and Elliptic Orbits

The quantization of angular momentum proposed by Bohr was generalized by Sommerfeld and Wilson with the aim to apply it to atoms more complex than the hydrogen one [12].

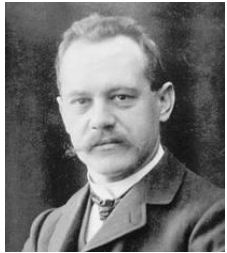

**Arnold Sommerfeld**

Supposing an atomic system with $g$ degrees of freedom represented by the generalized coordinates $x_1, \dots, x_q$, which are time-dependent, and by the conjugate momenta $p_1, \dots p_g$, Sommerfeld and Wilson proposed the following quantization rule:

$$\int p_1 dx_1 = l_1 h, \dots \dots \dots, \int p_g dx_g = l_g h \qquad (\mathbf{3}.\boldsymbol{a})$$

where the integrals are calculated over the range of periodicity of the correspondent variable. The integrals (**3.a**) are just the mechanical actions of the electron. Sommerfeld and Wilson supposed that for a given value of the principal quantum number $n$, to which corresponds a single energy, exist $n$ possible orbital paths characterized by the new quantum numbers $l_1, \dots l_g$, called secondary or azimuthal quantum numbers. These paths are ellipsis of different eccentricity.

Is important highlight that orbits having the same principal quantum number and different azimuthal numbers have the same energy and they differ only for their geometrical orientation! While in the circular orbit the electron is always equidistant from the nucleus, in the elliptical ones the distance changes in a periodical way according to its eccentricity. Such a variation leads to a continuous modification of the electron velocity that increases as the electron approaches to the core. Supposing a relativistic behavior of the electron, the mass $m = m_0 \gamma$ depends on eccentricity of the elliptical orbit. Since the energy is quantized and depends on the mass $m_e$ it follows that also the energy of the elliptical orbits changes with the eccentricity! Let me write this concept in the mathematical formalism. To such a purpose, the relativistic form of the total energy of the atom is given by the sum of the relativistic version of the kinetic energy **(2.c)** and coulombian potential energy:

$$E = \gamma m_0 c^2 - m_0 c^2 - \frac{1}{4\pi\varepsilon_0} Z q_e^2 = m_0 c^2 (\gamma - 1) - \frac{1}{4\pi\varepsilon_0} \frac{1}{r} Z q_e^2$$

where Z is the atomic number ($Z q_e$ is the nucleus charge) and $m_0$ the rest mass of electron. Setting the new variable $u = (1/r)$ we get:

$$\gamma = 1 + \frac{E}{m_0 c^2} + \frac{1}{4\pi\varepsilon_0} \frac{u Z q_e^2}{m_0 c^2} \qquad (\boldsymbol{3.b})$$

Recalling the theory of the motion of a particle in a central field, we can perform a changing of variables writing the component of the angular momentum in terms of polar coordinates (the central motion is flat):

$$r = (x, y) = (\rho \cos\theta, \rho \sin\theta) \quad \rightarrow \quad p_\rho = m\dot{\rho} \qquad p_\theta = m\rho^2 \dot{\theta}$$

Sommerfeld supposed that both angular momentum and energy was conserved, which means $p_\rho$, $p_\theta$ and E are constants of motion. We calculate now the ratio between the two polar components of angular momentum:

$$\frac{p_\rho}{p_\theta} = \frac{m\dot{\rho}}{m\rho^2 \dot{\theta}} = \frac{\dot{\rho}}{\rho^2 \dot{\theta}} = \frac{1}{\rho^2} \frac{d\rho}{d\theta} = -\frac{d(1/\rho)}{d\theta} \quad \Rightarrow \quad p_\rho = p_\theta \frac{1}{\rho^2} \frac{d\rho}{d\theta}$$

Recalling the quantization rule of Sommerfeld and Wilson:

$$\oint p_\rho d\rho = n_\rho h \qquad \oint p_\theta d\theta = n_\theta h$$

and applying the Binet formula we can write the equation of motion of relativistic electron:

$$\frac{d^2(1/\rho)}{d\theta^2} + \gamma^2 \left[ \frac{1}{\rho} - \frac{m_0 Z q_e^2}{p_\theta^2 \gamma^2} \left( 1 + \frac{E}{m_0 c^2} \right) \right] = 0 \qquad (\boldsymbol{3.c})$$

where the coefficient $\gamma$ is $\left(1 - (v^2/c^2)\right)^{-1/2}$. Solving the second order differential equation we get the law of motion:

$$\rho(\theta) = \frac{p_\theta^2 \gamma^2}{m_0 Z q_e^2} \left( 1 + \frac{E}{m_0 c^2} \right)^{-1} \frac{1}{1 + A \cos \gamma\theta}$$

where A is an integration constant to be calculated setting the boundaries conditions. We have now all the elements to arrive to the final result that is represented by the following relativistic equation (all mathematical details are found on the original article [12]):

$$\frac{E}{m_0 c^2} = \left(1 + \frac{\alpha^2 Z^2}{\left(n_\rho + \sqrt{n_\theta^2 - Z^2 \alpha^2}\right)^2}\right)^{-1/2} - 1 \qquad (\boldsymbol{3.d})$$

where $\alpha$ is the fine-structure constant given by:

$$\alpha = \frac{q_e^2}{4\pi\varepsilon_0 c\hbar}$$

whose meaning is of capital importance in quantum mechanics and will be discussed further. The term $n_\rho$ in (**3.d**) is the principal quantum number while $n_\theta$ is the secondary or angular one. The term under square root is just nothing that the relativistic coefficient $\gamma$ which, as expected on the basis of the discussion made above, depends on the eccentricity of the orbit and it's related to the secondary quantum number. In particular, increasing the quantum numbers $n_\rho$ and $n_\theta$ the electron tends to a non-relativistic behavior. We note that (**3.d**) has physical meaning only when $n_\theta^2 - Z^2\alpha^2 \gg 0$, when the angular momenta is greater than a minimum value! To a given quantum number $n_\rho$ are associated the values of $n_\theta$ according the rule $n_\theta = n_\rho - 1, n_\rho - 2, \dots, 0$. Thus, the relativistic theory of Sommerfeld predicts the fine structure of the atomic spectrum, according to the usual formula:

$$E(n, l) - E(n', l') = h\nu$$

When Sommerfeld and Wilson published their work the physicists of the period remained very fascinated and excited by the efficacy of their theory whose correctness has helped to confirm the theory of special relativity. However, the fine structure of hydrogen spectra shows that the lines have different intensities whose magnitude cannot be predicted by the Sommerfeld theory. Experiments carried out after Sommerfeld publication (1916) showed that the fine structure is due to the spin-orbit coupling (a phenomenon not explainable by the Bohr-Sommerfeld theory). Moreover, the spacing between some lines was not at all in agreement with the theoretical results. The improvement of the experimental techniques and the designing of instruments more and more accurate led the Sommerfeld theory fragile till to the born of the modern quantum mechanics due to Eisenberg, Schrodinger, Dirac and others. In the literature often the theories of Bohr and Sommerfeld are recalled as the *Old Quantum Mechanics*, but without any doubt they represent the beginning of a scientific period full of new and revolutionary ideas that opened the way to the new quantum physics, facilitating the work done by Schrodinger and colleagues.

## 4. The Wave Behavior of Matter

At the beginning of XX century physicists discovered that light and matter can behave like waves or particles depending on the nature of the performed experiment (photoelectric effect, double slit experiment, Compton effect, etc.). Soon, theoretical physicist began developing a theory able to explain these new and unexpected phenomena. Luis De Broglie, in its bachelor thesis, had the intuition to highlight the parallelism between the equations of electromagnetic waves and those of material particle motion [9,13-14]. He was so able to formulate a mathematical formalism where matter can be studied using the wave equations!

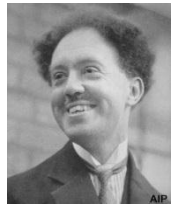

**Luis De Broglie**

Let's consider a monochromatic electromagnetic wave propagating in the vacuum in the direction given by the wave vector $\boldsymbol{k}$. The wave front is plane and perpendicular to the wave vector $\boldsymbol{k}$; its equation is given by:

$$\boldsymbol{k} \cdot \boldsymbol{r} = const. \qquad (\boldsymbol{4.a})$$

On the other words, the wave front is the locus of points having the same phase. From the electromagnetism such a wave is described by the following function:

$$\psi(\boldsymbol{r},t) = A_0 e^{\{i(\boldsymbol{k}\cdot\boldsymbol{r}-\omega t)\}} \qquad (\boldsymbol{4.b})$$

where $\omega = 2\pi\nu$ is the angular frequency. $A_0$ is the maximum amplitude of the wave and its squared modulus $|A_0|^2$ represents its intensity. Over time the wave front moves along the direction of the vector $\boldsymbol{k}$ in concordance of phase according to the equation:

$$\boldsymbol{k} \cdot \boldsymbol{r} - \omega t = const. \qquad (\boldsymbol{4.c})$$

All the consecutive wave fronts are equidistant by a length $\lambda = (2\pi/k)$ and move with a phase velocity given by:

$$v_f = \lambda\nu = \frac{\omega}{k} \qquad (\boldsymbol{4.d})$$

that coincides with the speed of light if wave travel in vacuum. In the case the wave propagates in an isotropic material medium the phase velocity becomes $v_g = (c/n)$ where $n$ is refraction index. If the material is anisotropic then the refraction index changes from point to point, which means that also the phase velocity is not more constant and the geometry of wave front deviates from planarity. In these cases the wave front equation becomes:

$$\psi_0(\boldsymbol{r}) - \omega t = const. \qquad (\boldsymbol{4.e})$$

and the wave function will be the following:

$$\psi(\boldsymbol{r}, t) = A_0 e^{\{i(\psi_0(\boldsymbol{r}) - \omega t)\}}$$

Substituting the last one in the D'Alambert equation:

$$\nabla^2 \psi(\boldsymbol{r}, t) - \frac{1}{v_f^2} \frac{\partial^2 \psi(\boldsymbol{r}, t)}{\partial t^2} = 0$$

we get:

$$-A_0 \big(\nabla \psi_0(\boldsymbol{r})\big)^2 e^{\{i(\psi_0(\boldsymbol{r}) - \omega t)\}} + \frac{\omega^2}{v_f^2} A_0 e^{\{i(\psi_0(\boldsymbol{r}) - \omega t)\}} = 0$$

Eliminating the complex exponential, which never vanishes, we get the equation:

$$\big(\nabla \psi_0(\boldsymbol{r})\big)^2 + \frac{\omega^2}{v_f^2} = 0 \qquad (\boldsymbol{4.f})$$

The phase velocity in whatever point is given by:

$$v_f(\boldsymbol{r}) = \frac{c}{n(\boldsymbol{r})} = \frac{\lambda_0 \nu}{n(\boldsymbol{r})} = \frac{2\pi \nu}{n(\boldsymbol{r}) \frac{2\pi}{\lambda_0}} = \frac{\omega}{n(\boldsymbol{r}) k_0}$$

and replacing it in (**4.f**) it's obtained:

$$-\big(\nabla \psi_0(\boldsymbol{r})\big)^2 + \omega^2 \frac{n^2 k_0{}^2}{\omega^2} = 0 \quad \rightarrow \quad \frac{1}{k_0{}^2} \big(\nabla \psi_0(\boldsymbol{r})\big)^2 = n(\boldsymbol{r})^2 \qquad (\boldsymbol{4.g})$$

where $k_0$ is the modulus of wave vector in the vacuum. If the wave is polychromatic then every component will satisfy to the one's own D'Alambert equation, travelling in the medium with the one's own phase velocity. In that case we define group velocity the following quantity:

$$v_g = \frac{d\omega}{dk} \qquad (\boldsymbol{4.h})$$

Let's consider now a material particle with mass $m_0$ moving at the velocity $\boldsymbol{v}$ in a conservative field characterized by the potential $V(\boldsymbol{r})$. According to the Hamiltonian mechanics the total energy E of the particle is given by:

$$E = \frac{p^2}{2m_0} + V(\boldsymbol{r})$$

The particle action, as Lagrangian integral, is:

$$S(\boldsymbol{r}, t) = \int L(q_1, \dots, q_n, \dot{q}_1, \dots, \dot{q}_n, t) dt$$

Since the Lagrangian function is given by the difference between the kinetic T and potential V energies, the action will assume the following form:

$$S(\boldsymbol{r}, t) = \boldsymbol{p} \cdot \boldsymbol{r} - Et \qquad (\boldsymbol{4.i})$$

This equation is parallel to the **(4.c)** under the hypothesis that $\boldsymbol{p} \equiv \boldsymbol{k}$ and $E \equiv \omega$, where the symbol $\equiv$ means analogous by mathematical symmetry. Applying the differential operator $\nabla$ to the function $S(\boldsymbol{r}, t)$ and calculating its squared it's obtained:

$$[\nabla S(\boldsymbol{r}, t)]^2 = p^2 = 2m_0[E - V(\boldsymbol{r})] \qquad (\boldsymbol{4.l})$$

The **(4.l)** is parallel to the **(4.g)** with $p^2 \equiv n^2$, which means the squared of linear momentum is analogous by mathematical symmetry to the squared of refraction index. From this parallelism we can write also the following relation:

$$[\nabla S(\boldsymbol{r}, t)]^2 = p^2 \equiv n^2 = \frac{1}{k_0{}^2}\left(\nabla \psi_0(\boldsymbol{r})\right)^2 \quad \rightarrow \quad p \equiv \frac{1}{k_0}\nabla \psi_0(\boldsymbol{r})$$

whose meaning will be clarified further. If the action $S(\boldsymbol{r}, t)$ is constant then we get:

$$S(\boldsymbol{r}, t) = \boldsymbol{p} \cdot \boldsymbol{r} - Et = const.$$

which is the equation of the plane analogous to the **(4.c)**. The velocity of the material particle can be assimilated to that of an electromagnetic wave according to the following equality:

$$v_f = \frac{\omega}{k} \equiv \frac{E}{p} \qquad (\boldsymbol{4.m})$$

The **(4.m)** resumes the whole parallelism between the geometrical optics of waves and the dynamics of material particles. On the basis of this equality De Broglie supposed that a particle could be considered like a material wave with energy $E = h\nu$ and propagating with a phase velocity $v_f = \lambda\nu$. Recalling the **(4.m)**:

$$v_f = \frac{E}{p} = \frac{h\nu}{p} \quad \Rightarrow \quad \lambda = \frac{h}{p} \qquad (\boldsymbol{4.n})$$

According to the De Broglie hypothesis is possible to associate to whatever particle a wave with wavelength given by the **(4.n)**. From the Hamiltonian function the linear momentum is:

$$p = \sqrt{2m_0(E - V)}$$

so that the wavelength of the material wave becomes:

$$\lambda = \frac{h}{\sqrt{2m_0(E - V)}} \qquad (\boldsymbol{4.o})$$

Recalling always the parallelism between geometrical optics and particle dynamics, in analogy with the **(4.b)** the function of the material wave can be written as:

$$\psi(\boldsymbol{r}, t) = A_0 e^{\{i(\boldsymbol{p} \cdot \boldsymbol{r} - Et)\}} \qquad (\boldsymbol{4.p})$$

Let's suppose the particle it's moving with relativistic velocity $v$; according to Einstein theory it's energy is:

$$E = c\sqrt{p^2 + m_0^2 c^2}$$

Replacing this expression in the (**4.m**) we get:

$$v_f = \frac{E}{p} = c\sqrt{1 + \frac{m_0^2 c^2}{p^2}} = c\sqrt{1 + \frac{m_0^2 c^2}{m_0^2 v^2}} = c\sqrt{1 + \frac{c^2}{v^2}}$$

The term under square root is always greater than 1, which means that the phase velocity of the material wave is greyer than that of light! Such a result is meaningless and De Broglie was able to solve the problem assuming that the particle velocity is not correlated to that of the travelling wave. When the motion is relativistic we need to consider the particle as a group of waves, usually called wave packet. In analogy with the (**4.h**) the group velocity becomes:

$$v_g = \frac{d\omega}{dk} = \frac{dE}{dp} = \frac{c}{\sqrt{1 + \frac{c^2}{v^2}}}$$

which is always smaller than that of light. The wavelength of the relativistic particle is:

$$\lambda = \frac{h}{p} = \frac{hc}{\sqrt{E^2 - m_0^2 c^4}} \qquad (\boldsymbol{4.q})$$

We note the De Broglie wavelength (**4.n**) is equal to that obtained by Bohr in the (**2.d**) postulating the quantization of the angular momentum. The merit of De Broglie was that to rationalize from the physical point of view the issue obtained by Bohr.

Starting from (**2.b**) and (**2.g**) is possible to obtain also the electron velocity in the hydrogen atom:

$$v = \frac{q_e^2}{4\pi\varepsilon_0 n\hbar} = \frac{2188}{n} \quad Km/s$$

So, the electron velocity in the hydrogen atom is always less than 1% of that of light; that means we may consider the hydrogen electron like a non-relativistic particle. However, as will be proven further, in spite of its velocity the formulation of a relativist equation will lead to results able to explain, without the need of any further postulate, the physical reality of electron getting more robust and complete the Schrodinger quantum theory.

## 5. The Schrodinger Equation

In 1926 Erwin Schrödinger, using the De Broglie hypothesis, after few months of deep work formulated the homonymous equation that still represent the main tool of the whole non relativistic quantum mechanics [15-18].

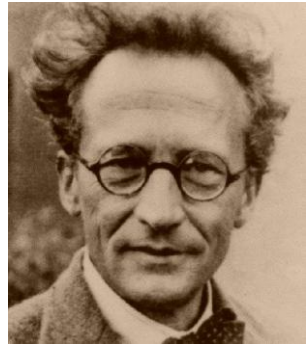

**Erwin Schrodinger**

Schrödinger considered the electron like a standing wave where the spatial and time parts are independents and separable:

$$\varphi(\boldsymbol{r}, t) = \psi(\boldsymbol{r}) cos\omega t \qquad (\boldsymbol{5.a})$$

Here $\omega$ is the angular frequency given by $2\pi\nu$, while $\nu$ is the frequency of material wave that, according to the De Broglie formula, is given by $\nu = v/\lambda = vh/p$. The hypothesis of the atomic electron as standing wave is sustainable because the experimental results show that its energy is conserved. Putting the **(5.a)** in the D'Alambert equation we get:

$$\nabla^2\varphi(\boldsymbol{r}, t) = \nabla^2\psi(\boldsymbol{r}) cos\omega t$$

$$\frac{\partial^2\varphi(\boldsymbol{r}, t)}{\partial t^2} = -\psi(\boldsymbol{r})\omega^2 cos\omega t$$

and so the equation becomes:

$$\nabla^2\psi(\boldsymbol{r}) cos\omega t = -\frac{\omega^2}{v^2}\psi(\boldsymbol{r}) cos\omega t$$

$$\nabla^2\psi(\boldsymbol{r}) = -\frac{4\pi^2}{\lambda^2}\psi(\boldsymbol{r}) \qquad (\boldsymbol{5.b})$$

The **(5.b)** is a second order differential equation where $\psi(\boldsymbol{r})$ is eigenfunction of the laplacian operator with eigenvalue $-4\pi^2/\lambda^2$. This last is the squared of the wave vector modulus:

$$\frac{4\pi^2}{\lambda^2} = k^2$$

Supposing that the particle moves with a non-relativistic velocity in a field with potential energy $V$, then according to the De Broglie theory its wavelength is given by:

$$\lambda = \frac{h}{p} = \frac{h}{\sqrt{2m_0(E - V)}}$$

that substituted in **(5.b)** gives:

$$\nabla^2\psi(\boldsymbol{r}) = -\frac{4\pi^2 2m_0(E - V)}{h^2}\psi(\boldsymbol{r}) = -\frac{8\pi^2 m_0 E}{h^2}\psi(\boldsymbol{r}) + \frac{8\pi^2 m_0 V}{h^2}\psi(\boldsymbol{r})$$

$$-\frac{h^2}{8\pi^2 m_0}\nabla^2\psi(\boldsymbol{r}) + V\psi(\boldsymbol{r}) = E\psi(\boldsymbol{r}) \qquad (\boldsymbol{5.c})$$

The **(5.c)** is the Schrödinger eigenvalue equation for a non-relativistic particle whose behavior is that of a standing material wave. Comparing the Hamilton function of the classical mechanics $E = T + V$ with the Schrödinger equation we can state the following mathematical equivalences:

$$T_H \equiv \left(-\frac{h^2}{8\pi^2 m_0}\nabla^2\right)_S$$

$$V_H \equiv V_S$$

where the under scripts H and S refer to Hamilton and Schrödinger. So, the classical kinetic energy is related to a quantum differential operator while the potential $V_H$ corresponds to an operator coincident with its classical function. Using the first of these equivalences it's easy to show another and more important relation between the classical momentum and the quantum differential operator:

$$T = \frac{p^2}{2m_0} \equiv -\frac{h^2}{8\pi^2 m_0}\nabla^2 \;\Rightarrow\; p \equiv i\hbar\nabla = -\frac{\hbar}{i}\nabla \qquad (\boldsymbol{5.d})$$

We note that the sign of the quantum momentum may be without any distinction positive or negative (in the **(5.d)** we wrote it with positive sign according the great part of the literature). This sign is an operatorial *memory* arising from the possible orientations of the classical vector $\boldsymbol{p}$. To avoid misunderstanding we will use a letter with a tilde to label a quantum operator and a letter without any sign to label the correspondent classical quantity. By these premises, following the Hamilton picture where the main physical quantities are formulated on the base of impulse $\boldsymbol{p}$, position $\boldsymbol{r}$, mass $m$ and time $t$, we can assume an analogous quantum mechanics picture, where the impulse is replaced by the operator **(5.d)** and the quantities $\boldsymbol{r}$, $m$ and $t$ remain equals to the classical ones. This assumption, coming from the comparison between equation **(5.c)** and the Hamilton function, will be one of the physical-mathematical postulates of the whole quantum mechanics theory.

The Schrödinger equation doesn't contain any information about time evolution of the material standing wave, neither in the operatorial terms nor in the wave function. About that let's consider the velocity of the material wave given by the **(5.m)**:

$$v_f = \frac{E}{p} \quad \rightarrow \quad E = pv_f$$

Here $v_f$ is the particle velocity and we are still in the classical theory. Passing to the formalism of quantum mechanics we can write the energy operator as follows:

$$\hat{E} = i\hbar \frac{\partial}{\partial r} \frac{\partial r}{\partial t} = i\hbar \frac{\partial}{\partial t} \qquad (\mathbf{5.e})$$

So, we found two ways of writing the total energy of the material wave in terms of quantum operators: that given by (**5.e**), which contains only the time, and that given by the sum of the operators $T_s$ and $V_S$:

$$\hat{E} = -\frac{h^2}{8\pi^2 m_0} \nabla^2 + V \qquad (\mathbf{5.e'})$$

which contains only the spatial variable. Because of the equality (**5.e**) and (**5.e'**) the Schrödinger equation (**5.c**) can be rewritten as:

$$\left[ -\frac{h^2}{8\pi^2 m_0} \nabla^2 + V \right] \varphi(\boldsymbol{r}, t) = i\hbar \frac{\partial}{\partial t} \varphi(\boldsymbol{r}, t) \qquad (\mathbf{5.f})$$

which is known as the time depending Schrödinger equation. Since the operator $\hat{E}$ has the same physical meaning of the classical Hamiltonian function (total energy of the system), for convenience we write it by the capital letter H. It's a linear and hermitian differential operator with real eigenvalues [21]. This property proves the physical correctness of equation (**5.c**) being the energy a real number just like all the observables! It's easy proving that also the linear momentum operator is hermitian:

$$\hat{p} = i\hbar \frac{\partial}{\partial r} = -i\hbar \left( -\frac{\partial}{\partial r} \right) = (i\hbar)^\dagger \left( \frac{\partial}{\partial r} \right)^\dagger = \hat{p}^\dagger$$

$\hat{p}$ is self-adjoint and so hermitian. Obviously also the quantities $m$, $\boldsymbol{r}$ and $t$ being reals are hermitian operators. We conclude that in quantum mechanics every physical quantity is expressed by a hermitian operator [21].

The eigenfunction $\varphi(\boldsymbol{r}, t)$ is a complex function representing the state of the system; its physical meaning will be explained further.

The Schrödinger approach is based on the assumption to consider the particle as a material wave that satisfies the D'Alambert equation; for that reason is defined as wave mechanics. Parallel to this picture is that of Heisenberg, which will be introduced further, based on a different mathematical formulation. The Schrödinger picture is the most used even if that of Heisenberg is more elegant concerning its mathematical formalism.

Schrödinger applied equation (**5.c**) to the hydrogen atom obtained the first quantum theory able to explain the experimental results, without the need to introduce ad hoc any other hypothesis different from that of De Broglie.

## 6. The Hydrogen Atom in the Picture of Wave Mechanics

Let be respectively $\boldsymbol{r}_N$ and $\boldsymbol{r}_e$ the position vectors of core and electron and $m_N$ and $m_e$ their mass. The framework is centered in the atomic center of mass. The reduced mass of the atom is $\mu = m_N m_e / (m_N + m_e)$. The Schrödinger equation can be written setting the Hamiltonian operator as follows [19]:

$$H = -\frac{\hbar^2}{2m_N}\nabla_N^2 - \frac{\hbar^2}{2m_e}\nabla_e^2 + V(\boldsymbol{r}_e - \boldsymbol{r}_N)$$

where the subscripts *N* and *e* are referred to the nucleus and electron. It is convenient rewrite the Hamiltonian using the coordinates of the center of mass and the reduced mass:

$$H = -\frac{\hbar^2}{2m_N + m_e}\nabla_{CM}^2 - \frac{\hbar^2}{2\mu}\nabla^2 + V(\boldsymbol{r}) \qquad (6.a)$$

where the subscript *CM* denotes the center of mass. The first term of the **(6.a)** is the kinetic energy operator of the center of mass, the second represents the kinetic energy of the reduced mass and the last one is the potential energy due to the electrostatic interaction between electron and core. Using the operator **(6.a)** the Schrödinger equation becomes:

$$H\varphi(\boldsymbol{r}_{CM}, \boldsymbol{r}) = E\varphi(\boldsymbol{r}_{CM}, \boldsymbol{r})$$

Since the motion of the center of mass is independent respect that of the reduced mass, the eigenfunction $\varphi(\boldsymbol{r}_{CM}, \boldsymbol{r})$ can be factorized as follows:

$$\varphi(\boldsymbol{r}_{CM}, \boldsymbol{r}) = \chi(\boldsymbol{r}_{CM})\psi(\boldsymbol{r}) \qquad (6.b)$$

The operator **(6.a)** can be rewritten as the sum of the Hamiltonian of the center of mass (first term of H) and the Hamiltonian of the reduced mass (second and third term of H):

$$H = H_{CM} + H_\mu$$

Applying the function **(6.b)** to this last operator we arrive to the following equation:

$$H\chi(\boldsymbol{r}_{CM})\psi(\boldsymbol{r}) = H_{CM}\chi(\boldsymbol{r}_{CM})\psi(\boldsymbol{r}) + H_\mu\chi(\boldsymbol{r}_{CM})\psi(\boldsymbol{r}) = E_{CM}\chi(\boldsymbol{r}_{CM})\psi(\boldsymbol{r}) + E_\mu\chi(\boldsymbol{r}_{CM})\psi(\boldsymbol{r}) =$$

$$= (E_{CM} + E_\mu)\chi(\boldsymbol{r}_{CM})\psi(\boldsymbol{r})$$

The Schrödinger equation for the hydrogen atom may be separated in two independent equations:

$$H_{CM}\chi(\boldsymbol{r}_{CM}) = E_{CM}\chi(\boldsymbol{r}_{CM}) \qquad (6.c)$$

$$H_\mu\psi(\boldsymbol{r}) = E_\mu\psi(\boldsymbol{r}) \qquad (6.d)$$

Let's to solve the equation **(6.c)** obtaining eigenfunction and energy about the free motion of the atom. This equation must be written in the explicit form:

$$-\frac{\hbar^2}{2(m_N + m_e)}\nabla^2_{CM}\chi(\boldsymbol{r}_{CM}) = E_{CM}\chi(\boldsymbol{r}_{CM})$$

$$\nabla^2_{CM}\chi(\boldsymbol{r}_{CM}) = -\frac{2(m_N + m_e)}{\hbar^2}E_{CM}\chi(\boldsymbol{r}_{CM})$$

The solutions of this differential equation are:

$$\chi(\boldsymbol{r}_{CM}) = A\,exp\left\{i\sqrt{\frac{2(m_N + m_e)E_{CM}}{\hbar^2}}\,r\right\} + B\,exp\left\{-i\sqrt{\frac{2(m_N + m_e)E_{CM}}{\hbar^2}}\,r\right\}$$

Since the energy $E_{CM}$ is an unknown quantity we can avoid this problem using the **(4.o)** so to obtain:

$$2(m_N + m_e)E_{CM} = \frac{h^2}{\lambda^2_{CM}}$$

where $\lambda_{CM}$ is the De Broglie wavelength of the hydrogen atom; substituting this result in the last expression we get:

$$\chi(\boldsymbol{r}_{CM}) = A\,exp\left\{i\frac{h}{\lambda_{CM}\hbar}r\right\} + B\,exp\left\{-i\frac{h}{\lambda_{CM}\hbar}r\right\}$$

and recalling that the wave vector modulus is given by $k = 2\pi/\lambda$ we arrive to the final function:

$$\chi(\boldsymbol{r}_{CM}) = Ae^{ik\cdot r} + Be^{-ik\cdot r} \qquad (\boldsymbol{6.e})$$

The **(6.e)** is the searched solution and is a typical plane wave that replaced in the equation **(6.c)** gives the total energy of the center of mass:

$$E_{CM} = \frac{\hbar^2 k^2}{2(m_N + m_e)}$$

The calculation of the two numerical constants A and B of the **(6.e)** can be done using the boundary conditions that, at the moment, is premature to set because we did not still explained the physical meaning of eigenfunction (probabilistic hypothesis of Born). Since there are not any restrictions on the choice of the wave vector modulus it follows that the energy $E_{CM}$ is not quantized, as expected for a free particle moving in an unlimited space.

Let's consider now the Schrödinger equation of the reduced mass:

$$-\frac{\hbar^2}{2\mu}\nabla^2\psi(\boldsymbol{r}) + V\psi(\boldsymbol{r}) = E_\mu\psi(\boldsymbol{r})$$

As well as made for the Bohr model of hydrogen atom, because of $m_N \gg m_e$ the reduced mass may be approximated to that of the electron. The Schrödinger equation then becomes:

$$-\frac{\hbar^2}{2m_e}\nabla^2\psi(\boldsymbol{r}) + V\psi(\boldsymbol{r}) = E\psi(\boldsymbol{r}) \qquad (\boldsymbol{6.f})$$

E is the total energy of the electron. The solution of this differential equation in Cartesian coordinates is quite difficult and can be considerably simplified using a more appropriate set of coordinates which reflect the symmetry of the physical system. The electron motion around the nucleus is equivalent to that of a particle in a central field with spherical symmetry. The set of coordinates we are seeking is the following:

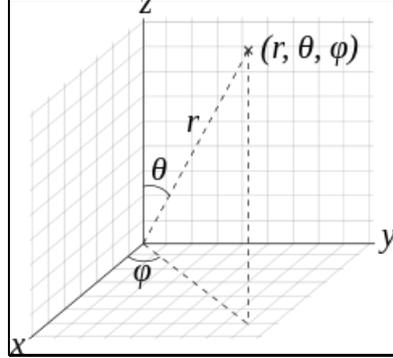

$$\begin{cases} x = r \sin \theta \cos \varphi \\ y = r \sin \vartheta \sin \varphi \\ z = r \cos \theta \end{cases} \qquad (\textbf{6.g})$$

To write the equation (**6.f**) in spherical coordinates is necessary to perform first a change of variables to the operator $\nabla^2$:

$$-\frac{\hbar^2}{2m_e}\left[\frac{1}{r^2}\frac{\partial}{\partial r}\left(r^2\frac{\partial}{\partial r}\right) + \frac{1}{r\sin^2\theta}\frac{\partial}{\partial \theta}\left(\sin\theta\frac{\partial}{\partial \theta}\right) + \frac{1}{r^2\sin^2\theta}\frac{\partial^2}{\partial \varphi^2}\right]\psi(r,\theta,\varphi) +$$

$$+V(r)\psi(r,\theta,\varphi) = E\psi(r,\theta,\varphi) \qquad (\textbf{6.h})$$

Since the coordinates $r, \theta, \varphi$ (generalized coordinates) are independent, the eigenfunction $\psi(r,\theta,\varphi)$ may be factorized in the product of one-variable functions:

$$\psi(r,\theta,\varphi) = R(r)\Theta(\theta)\Phi(\varphi)$$

The electrostatic energy $V(r)$ depends only on generalized coordinate $r$; according to the quantum mechanics rules discussed in the previous section the operator corresponding to this quantity is:

$$V(r) = -\frac{1}{4\pi\varepsilon_0}\frac{e^2}{r}$$

which is the typical expression of the potential energy of a field with central symmetry produced by two electrical charges with opposite sign. Replacing the factorized eigenfunction and the explicit form of the electrostatic potential energy in equation (**6.h**) and multiplying both members times $r^2\sin\theta/R(r)\Theta(\theta)\Phi(\varphi)$ we obtain the following differential equation with separable variables:

$$\frac{\sin^2\theta}{R(r)}\frac{\partial}{\partial r}\left(r^2\frac{\partial R}{\partial r}\right) + \frac{\sin\theta}{\Theta(\theta)}\frac{\partial}{\partial \theta}\left(\sin\theta\frac{\partial \Theta}{\partial \theta}\right) + \frac{2m_e}{\hbar^2}r^2\sin^2\theta\left(E + \frac{e^2}{4\pi\varepsilon_0 r}\right) = -\frac{1}{\Phi(\varphi)}\frac{\partial^2\Phi}{\partial \varphi^2}$$

The first member of this equation depends on the generalized coordinates $r, \theta$ while the second one is a function of the coordinate $\varphi$. The two members can be set equal to a numerical constant $m^2$. Working on the second member we obtain:

$$-\frac{1}{\Phi(\varphi)}\frac{\partial^2 \Phi}{\partial \varphi^2} = m^2 \quad \Rightarrow \quad \frac{\partial^2 \Phi}{\partial \varphi^2} = -m^2 \Phi(\varphi)$$

This is a simple eigenvalue differential equation of the second order whose solutions are:

$$\Phi_m(\varphi) = e^{\pm im\varphi}$$

These analytical functions must be monodrome, which means that for each $\varphi$ they can assume only a single value (that condition comes from the probabilistic interpretation of eigenfunction given by Born and will be discussed further), that leads to:

$$\Phi_m(\varphi) = \Phi_m(\varphi + 2\pi) = e^{\pm im(\varphi + 2\pi)} = e^{\pm im\varphi} + e^{\pm im2\pi}$$

The last equality implies that the constant $m$ must satisfy the following condition:

$$m = 0, \pm 1, \pm 2, \ldots, \pm n$$

Let's solve now the equation at the first member setting the second one equal to the constant $m^2$. This equation can be reworked to separate the terms depending on the variables $r$ and $\theta$:

$$\frac{1}{R(r)}\frac{\partial}{\partial r}\left(r^2 \frac{\partial R}{\partial r}\right) + \frac{2m_e}{\hbar^2}r^2\left(E + \frac{e^2}{4\pi\varepsilon_0 r}\right) = -\frac{1}{\Theta(\theta)sin\theta}\frac{\partial}{\partial \theta}\left(sin\theta \frac{\partial \Theta}{\partial \theta}\right) + \frac{m^2}{sin^2\theta}$$

Since each member of the equation depend only on a single variable they must be set equals to a same numerical constant that, for mathematical convenience, we will denote by $l(l+1)$. The second member then becomes:

$$\frac{1}{sin\theta}\frac{\partial}{\partial \theta}\left(sin\theta \frac{\partial \Theta}{\partial \theta}\right) - \frac{m^2 \Theta(\theta)}{sin^2\theta} + l(l+1) = 0$$

If $l$ is a natural number then the written equation is that of Legendre. Supposing $m = 0$ we have:

$$\frac{1}{sin\theta}\frac{\partial}{\partial \theta}\left(sin\theta \frac{\partial \Theta}{\partial \theta}\right) + l(l+1) = 0$$

whose solution are Legendre polynomials:

$$\Theta_l(cos\theta) = \frac{1}{2^l l!}\frac{d(cos^2\theta - 1)^l}{d(cos\theta)^l}$$

Otherwise, if $m \neq 0$ then the solutions are the Legendre functions:

$$\Theta_l^m(\theta) = sin^{|m|}\theta \frac{d^{|m|}}{d(cos\theta)^{|m|}}\left[\frac{1}{2^l l!}\frac{d(cos^2\theta - 1)^l}{d(cos\theta)^l}\right]$$

It must be noting that if the constant was different from $l(l+1)$, with $l$ natural number, then the solutions would not have the periodicity of $2\pi$.

Finally, let's consider the equation of the first member setting it equal to the usual constant $l(l+1)$:

$$\frac{1}{r^2}\frac{\partial}{\partial r}\left(r^2\frac{\partial R}{\partial r}\right)+\frac{2m_e}{\hbar^2}r^2\left(E+\frac{e^2}{4\pi\varepsilon_0 r}\right)R(r)-\frac{l(l+1)}{r^2}R(r)=0 \qquad (\boldsymbol{6.i})$$

This is the Laguerre differential equation whose solutions are:

$$R_{nl}(r)=\sqrt{\left(\frac{2m_e e^2}{\hbar^2}\right)^3\frac{(n-l+1)!}{[2n(n+l)!]^3}e^{-\rho/2}\rho^l L_{n+l}^{2l+1}(\rho)}$$

where:

$$\rho=\frac{2m_e e^2}{n\hbar^2 4\pi\varepsilon_0}r \qquad \forall n\epsilon N-\{0\} \qquad (\boldsymbol{6.i'})$$

$$L_{n+l}^{2l+1}(\rho)=\frac{d^{2l+1}}{d\rho^{2l+1}}L_{n+l}(\rho)$$

$$L_{n+l}(\rho)=e^{\rho}\frac{d^{n+l}}{d\rho^{n+l}}(\rho^{n+l}e^{-\rho})$$

Introducing the function $R_{nl}(r)$ in the Laguerre equation (**6.i**) we get the value of the electron energy:

$$E_n=-\frac{m_e e^4}{2(4\pi\varepsilon_0)^2}\frac{1}{n^2\hbar^2} \qquad (\boldsymbol{6.l})$$

This result coincides to that (**2.f**) obtained by Bohr! That means the quantization of the angular momentum arbitrary done by Bohr agree with the quantum wave mechanics. Moreover, the constant term of the function (**6.i'**) is just the radius of the Bohr orbit (**6.g**):

$$a_B=\frac{n^2\hbar^2 4\pi\varepsilon_0}{m_e e^2}$$

The integer numbers $n,l,m$ play a fundamental role in the physical chemistry field. They are defined as follows:

- $n$= main quantum number
- $l$= secondary or azimuthal or orbital quantum number
- $m$= magnetic quantum number

The first one gives, as shown by the (**6.h**), the total energy of the electron state (eigenfunction). Since $n$ can assume only positive integer numbers (different from zero) the electron energy of the hydrogen atom is quantized. The orbital quantum number, on the other hand, determines the geometrical shape of the eigenfunction; it coincides with that introduced by Sommerfeld to extend the Bohr model to atoms more complex than the hydrogen one. Since the wave function $\psi(r,\theta,\varphi)$ must satisfy appropriate mathematical conditions, arising from its probabilistic interpretation, the quantum number $l$ can assume only values given by $l \le n-1$. Finally, the magnetic quantum number $m$ is connected to the effects produced by the magnetic field due to the electron motion around the core. These effects can be seen acquiring the atomic spectra in an external magnetic field. Just like the orbital quantum number, the mathematical conditions forcing the eigenfunction to be physically acceptable determine the limit of $m$ that must range within $-l \le m \le l$.

Let's consider some examples of wave functions $\psi(r,\theta,\varphi) = R_{n,l}(r)\Theta_{l,m}(\theta)\Phi_m(\varphi)$ associated to a given triplet of quantum numbers $(n,l,m)$:

$$(n=1, l=0, m=0) \;\;\rightarrow\;\; \frac{1}{\sqrt{\pi}} a_B^{-\frac{3}{2}} e^{-r/a_B}$$

$$(n=2, l=0, m=0) \;\;\rightarrow\;\; \frac{1}{4\sqrt{2\pi}} a_B^{-\frac{3}{2}} \left(2 - \frac{r}{a_B}\right) e^{-r/2a_B}$$

$$(n=2, l=1, m=0) \;\;\rightarrow\;\; \frac{1}{4\sqrt{2\pi}} a_B^{-\frac{3}{2}} \frac{r}{a_B} cos\theta e^{-r/2a_B}$$

$$(n=2, l=1, m=1) \;\;\rightarrow\;\; \frac{1}{4\sqrt{2\pi}} a_B^{-\frac{3}{2}} \frac{r}{a_B} sin\theta e^{-r/2a_B} e^{i\varphi}$$

$$(n=2, l=1, m=-1) \;\;\rightarrow\;\; \frac{1}{4\sqrt{2\pi}} a_B^{-\frac{3}{2}} \frac{r}{a_B} sin\theta e^{-r/2a_B} e^{-i\varphi}$$

The function $R_{n,l}(r)$ is defined radial function while the product $\Theta_{l,m}(\theta)\Phi_m(\varphi)$ is defined spherical harmonic and usually in the literature is denoted by the symbol $Y_l^m(\theta,\varphi)$. It's an eigenfunction of an operator connected with the electron angular momentum and will be discussed in the next section. The eigenfunctions are called orbitals, in analogy with the orbits of the Bohr model, although they are not trajectories! For these orbitals are used the symbols listed in table 6:

| Orbital Quantum number | $l$ | 0 | 1 | 2 | 3 |
|---|---|---|---|---|---|
| Magnetic number | $m$ | $0, \pm 1$ | $0, \pm 1$ | $0, \pm 1, \pm 2$ | $0, \pm 1, \pm 2, \pm 3$ |
| Orbital Symbol | | $s$ | $p$ | $d$ | $f$ |

**Table 6**

In front of the orbital symbol is specified the value of the main quantum number:

- $n = 1 \implies 1s$

- $n = 2 \implies 2s, \ 2p \ (3 \ degenerate \ orbitals)$

- $n = 3 \implies 3s, \ 3p \ (3 \ degenerate \ orbitals), \ 3d \ (5 \ degenerate \ orbitals)$

The degeneration of the orbitals $p$, $d$ and $f$ is due to the fact that in formula **(6.l)** the quantum number $l$ and $m$ do not appear!

Concerning the shape of the radial functions we can state that:

1. for a given orbital number the corresponding function is zero at $n - 1$ (called also nodes);

2. all functions, independently from their orbital number, tend to zero increasing the distance from the nucleus. However, this trend gets slower increasing the value of the main quantum number;

3. only function $s$ has a pick in correspondence of the core while the other ones are zero.

To complete the solution of the Schrödinger equation for hydrogen atom the time-dependence of the eigenfunction has to be taken in account. To do this we need to solve equation **(6.f)**:

$$H_e \psi(r, \theta, \varphi, t) = i\hbar \frac{\partial}{\partial t} \psi(r, \theta, \varphi, t) = E\psi(r, \theta, \varphi, t)$$

where $H_e$ is the quantum mechanics Hamiltonian in spherical coordinates with the approximation $\mu \cong m_e$. The eigenfunction $\psi(r, \theta, \varphi, t)$ can be factorized as the product of the spatial orbital $\psi(r, \theta, \varphi)$ times a function $f(t)$ of time; the equation becomes:

$$i\hbar \frac{\partial}{\partial t} \psi(r, \theta, \varphi)f(t) = E\psi(r, \theta, \varphi)f(t) \quad \implies$$

$$\frac{\partial}{\partial t} f(t) = \frac{E}{i\hbar} f(t) = -i\frac{E}{\hbar} f(t) \qquad (\boldsymbol{4.4.m})$$

and the general solution is:

$$f(t) = e^{-iEt/\hbar} \qquad (\boldsymbol{6.n})$$

So, the full solution of equation **(6.i)** is:

$$\psi(r, \theta, \varphi) = R_{n,l}(r)\Theta_{l,m}(\theta)\Phi_m(\varphi) e^{-iEt/\hbar}$$

Replacing this function into equation **(6.h)** we obtain just the same energy given by the **(6.l)**. That means the total energy of electron does not depend on time and is therefore a constant of motion. The Schrödinger theory is then able to explain the stability of hydrogen atom without making use of any arbitrary assumption and give us the full explanation of its spectrum.

## 7. The Angular Momentum of Hydrogen Atom

Since the motion of the electron takes place around the core we have to find its angular momentum represented by the vector $\boldsymbol{L}$:

$$\boldsymbol{L} = \boldsymbol{r} \times \boldsymbol{p} = det\begin{pmatrix} \boldsymbol{i} & \boldsymbol{j} & \boldsymbol{k} \\ x & y & z \\ p_x & p_y & p_z \end{pmatrix} = \boldsymbol{i}\big(yp_z - zp_y\big) + \boldsymbol{j}\,\big(zp_x - xp_z\big) + \boldsymbol{k}(xp_y - yp_x)$$

The vectorial components are:

$$\begin{cases} L_x = yp_z - zp_y \\ L_y = zp_x - xp_z \\ L_z = xp_y - yp_x \end{cases} \qquad (\boldsymbol{7.a})$$

To obtain the quantum operator associated to each scalar component **(7.a)** we have to replace to $\boldsymbol{p}$ the operator $i\hbar\nabla$ and to $\boldsymbol{r}$ the same vector; the quantum version of **(7.a)** becomes:

$$\begin{cases} \hat{L}_x = i\hbar\left(y\dfrac{\partial}{\partial z} - z\dfrac{\partial}{\partial y}\right) \\[2mm] \hat{L}_y = i\hbar\left(z\dfrac{\partial}{\partial x} - x\dfrac{\partial}{\partial z}\right) \\[2mm] \hat{L}_z = i\hbar\left(x\dfrac{\partial}{\partial y} - y\dfrac{\partial}{\partial x}\right) \end{cases} \qquad (\boldsymbol{7.b})$$

In spherical coordinates the operators are:

$$\begin{cases} \hat{L}_x = i\hbar\left(-sin\varphi\,\dfrac{\partial}{\partial\theta} - ctg\theta cos\varphi\,\dfrac{\partial}{\partial\varphi}\right) \\[2mm] \hat{L}_y = i\hbar\left(cos\varphi\,\dfrac{\partial}{\partial\theta} - ctg\theta sin\varphi\,\dfrac{\partial}{\partial\varphi}\right) \\[2mm] \hat{L}_z = i\hbar\dfrac{\partial}{\partial\varphi} \end{cases}$$

Let's calculate now the squared operator $\hat{L}^2$ according to the scalar product rule:

$$\hat{L}^2 = -\hbar^2\left[\frac{1}{sin^2\theta}\frac{\partial}{\partial\theta}\left(sin\theta\frac{\partial}{\partial\theta}\right) + \frac{1}{sin^2\theta}\frac{\partial^2}{\partial\varphi^2}\right]$$

This operator is just the angular part of the Hamiltonian **(6.h)**. That means the differential equation $\hat{L}^2 Y(\theta,\varphi) = L^2 Y(\theta,\varphi)$ has the same solutions of the Schrödinger one:

$$Y(\theta,\varphi) = \Theta_{l,m}(\theta)\Phi_m(\varphi) \qquad (\boldsymbol{7.c})$$

The eigenvalues $L^2$ are given by $\hbar^2 l(l+1)$ and they are typical of the Legendre differential equations. The **(7.c)** is the spherical harmonics calculated in the previous section. It must be observed that all the operators $\hat{L}_x$, $\hat{L}_y$, $\hat{L}_z$ are hermitian and, as expected, have real eigenvalues.

## 8. Commutation Relations

The algebra of hermitian operators states that if two operators have the same eigenfunction then they commute and vice versa [21]. In the previous section we stressed that the operator $\hat{L}^2$ has the same eigenfunctions of the Hamiltonian $H_e$; that means the two operators commute $\left[H_e, \hat{L}^2\right] = 0$.

Let's now analyze the other possible commutator among the quantum operators obtained in the section 7:

$$[\hat{L}_x, \hat{L}_y] \;,\; [\hat{L}_y, \hat{L}_z] \;,\; [\hat{L}_x, \hat{L}_z] \;,\; [\hat{L}^2, \hat{L}_x] \;,\; [\hat{L}^2, \hat{L}_y] \;,\; [\hat{L}^2, \hat{L}_z]$$

At the beginning we calculate the first commutator using the explicit operatorial forms **(7.b)** and whatever continues function $f(x, y, z)$ derivable at least two times:

$$[\hat{L}_x, \hat{L}_y] f(x, y, z) = \hat{L}_x \hat{L}_y f(x, y, z) - \hat{L}_y \hat{L}_x f(x, y, z)$$

Calculating separately the two terms in the second member and using the Schwartz theorem we get:

$$\hat{L}_x \hat{L}_y f(x, y, z) = (i\hbar)^2 \left( y \frac{\partial}{\partial z} - z \frac{\partial}{\partial y} \right) \left( z \frac{\partial f}{\partial x} - x \frac{\partial f}{\partial z} \right) =$$

$$= (i\hbar)^2 \left( y \frac{\partial f}{\partial x} + yz \frac{\partial^2 f}{\partial z \partial x} - yx \frac{\partial^2 f}{\partial z^2} - z^2 \frac{\partial^2 f}{\partial y \partial x} + zx \frac{\partial^2 f}{\partial z \partial y} \right)$$

$$\hat{L}_y \hat{L}_x f(x, y, z) = (i\hbar)^2 \left( z \frac{\partial}{\partial x} - x \frac{\partial}{\partial z} \right) \left( y \frac{\partial f}{\partial z} - z \frac{\partial f}{\partial y} \right) =$$

$$= (i\hbar)^2 \left( zy \frac{\partial^2 f}{\partial x \partial z} - z^2 \frac{\partial^2 f}{\partial x \partial y} - xy \frac{\partial^2 f}{\partial z^2} + x \frac{\partial f}{\partial y} + xz \frac{\partial^2 f}{\partial z \partial y} \right)$$

Subtracting the two terms between them it's obtained:

$$[\hat{L}_x, \hat{L}_y] f(x, y, z) = (i\hbar) \left( y \frac{\partial f}{\partial x} - x \frac{\partial f}{\partial y} \right)$$

The terms within the bracket is just the explicit form of the operator $-\hat{L}_z$:

$$[\hat{L}_x, \hat{L}_y] = -i\hbar \hat{L}_z$$

We conclude that the commutator $[\hat{L}_x, \hat{L}_y]$ is non zero and the two operators $\hat{L}_x$ and $\hat{L}_y$ do not have the same eigenfunctions. In quantum mechanics it is usually to say that they have not simultaneous eigenstates. This property is of fundamental importance and is correlated to the Heisenberg uncertainty principle [22]. Like done for the first commutator we prove that:

$$[\hat{L}_y, \hat{L}_z] = -i\hbar \hat{L}_x$$

$$[\hat{L}_x, \hat{L}_z] = -i\hbar \hat{L}_y$$

The hermitian operators representing the components of the operator $\hat{L}$ do not commute and therefore they cannot have simultaneous eigenstates.

Let's consider now the commutator $[\hat{L}^2, \hat{L}_x]$ and, to simplify the calculus, we suppose a one-dimensional case:

$$\hat{L}_x = x i\hbar \frac{\partial}{\partial x} \;,\; \hat{L}^2 = \hat{L}_x \cdot \hat{L}_x = (i\hbar)^2 x \left( \frac{\partial}{\partial x} + x \frac{\partial^2}{\partial x^2} \right)$$

The explicit form of the commutator is $\left[\hat{L}^2, \hat{L}_x\right] = \hat{L}^2\hat{L}_x - \hat{L}_x\hat{L}^2$; the first term can be developed as:

$$\hat{L}^2\hat{L}_x = (i\hbar)^3\left(\frac{\partial}{\partial x} + x\frac{\partial^2}{\partial x^2}\right)\left(x\frac{\partial}{\partial x}\right) = (i\hbar)^3 x\left(\frac{\partial}{\partial x} + 3x\frac{\partial^2}{\partial x^2} + x^3\frac{\partial^3}{\partial x^3}\right)$$

while the second becomes:

$$\hat{L}_x\hat{L}^2 = (i\hbar)^3\left(x\frac{\partial}{\partial x}\right)\left(\frac{\partial}{\partial x} + x\frac{\partial^2}{\partial x^2}\right) = (i\hbar)^3 x\left(\frac{\partial}{\partial x} + 3x\frac{\partial^2}{\partial x^2} + x^3\frac{\partial^3}{\partial x^3}\right)$$

The two operatorial products are identical so that the whole commutator $\left[\hat{L}^2, \hat{L}_x\right]$ is zero. Following the same calculus just performed it's possible to prove that:

$$\left[\hat{L}^2, \hat{L}_x\right] = \left[\hat{L}^2, \hat{L}_y\right] = \left[\hat{L}^2, \hat{L}_z\right] = 0$$

All the components of the quantum angular momentum $\hat{L}$ commute with the operator $\hat{L}^2$ and they have the same eigenstates. If the commutation relations are satisfied then the number of differential equations to be solved may be reduced simplifying the study of the atomic system. For example, the solution of equation

$$\hat{L}^2 Y(\theta, \varphi) = \hbar^2 l(l+1) Y(\theta, \varphi)$$

gives simultaneously also the solutions of the differential equations associated to the components of the quantum angular momentum!

Let's calculate now the following commutators:

$$\left[x, \hat{p}_y\right] \ , \ \left[x, \hat{p}_z\right] \ , \ \left[y, \hat{p}_x\right] \ , \ \left[y, \hat{p}_z\right] \ , \ \left[z, \hat{p}_x\right] \ , \ \left[z, \hat{p}_y\right] \qquad (\mathbf{8.a})$$

Starting from the first and remembering the explicit form of linear impulse operator **(7.b)** we get:

$$\left[x, \hat{p}_y\right] = x\hat{p}_y - \hat{p}_y x = xi\hbar\frac{\partial}{\partial y} - i\hbar\frac{\partial}{\partial y}x = xi\hbar\frac{\partial}{\partial y} - xi\hbar\frac{\partial}{\partial y} = 0$$

This result may be easily proved applying the commutator to whatever continues function derivable at list one time:

$$\left[x, \hat{p}_y\right]f(x,y,z) = xi\hbar\frac{\partial f}{\partial y} - i\hbar\frac{\partial}{\partial y}[xf(x,y,z)] = xi\hbar\frac{\partial f}{\partial y} - f(x,y,z)i\hbar\frac{\partial x}{\partial y} - xi\hbar\frac{\partial f}{\partial y} = 0$$

where the term $f(x,y,z)i\hbar\frac{\partial x}{\partial y}$ is zero because the function $x$ is a constant respect the operator $\partial/\partial y$. Following the same procedure just performed for the commutator $\left[x, \hat{p}_y\right]$ we prove that all the commutators **(8.a)** are zero and the Cartesian coordinates of the position vector commute with the different Cartesian projections of the impulse operator. If the components of position and impulse vectors belong to the same Cartesian axis then we have:

$$\left[x, \hat{p}_x\right] \neq 0 \ , \ \left[y, \hat{p}_y\right] \neq 0 \ , \ \left[z, \hat{p}_z\right] \neq 0 \qquad (\mathbf{8.b})$$

which will lead us to the Heisenberg uncertainty principle. To prove these statements as usual we consider the first commutator applied to whatever function continues and derivable:

$$[x, \hat{p}_x]f(x) = xi\hbar \frac{\partial f}{\partial x} - i\hbar \frac{\partial}{\partial y}[xf(x)] = xi\hbar \frac{\partial f}{\partial x} - i\hbar f(x) - xi\hbar \frac{\partial f}{\partial x} = -i\hbar f(x)$$

We conclude that $[x, \hat{p}_y] = [y, \hat{p}_y] = [z, \hat{p}_z] = -i\hbar$ and position and impulse operators do not have the same eigenfunctions.

The considerations done about commutators recall the Poisson brackets concerning the Hamiltonian mechanics; this is another evidence of the deep connection between classical and quantum mechanics although their physical meaning are completely different. To such a purpose it's very easy proving that the properties of Poisson brackets (anti-symmetry, linearity, Leibnitz rule and Jacobi identity) are satisfied also by the quantum commutators:

$$[A, B] = -[B, A] \qquad (anti-symmetry)$$

$$[c_1 A + c_2 B, C] = c_1[A, C] + c_2[B, C] \qquad (linearity)$$

$$[AB, C] = A[B, C] + [A, C]B \qquad (Leibnitz\ rule)$$

$$\big[[A, B], C\big] + \big[[B, C], A\big] + \big[[C, A], B\big] = 0 \qquad (Jacobi\ identity)$$

The whole state of a quantum system is known when all differential equations associated to the commutators (usually said quantum conditions) are solved.

## 9. Probabilistic Interpretation of the Eigenfunction

The eigenfunctions of hydrogen atom are defined in $\mathbb{R}^3$ and their values are usually complex; changing the triplet of the quantum numbers $(n, l, m)$ they form an orthogonal base of the Hamiltonian operator. The physical meaning of these functions is difficult to explain even if, from the mathematical point of view, they lead to calculate the eigenvalues of the hermitian operator that are just the observables to which we are interested.

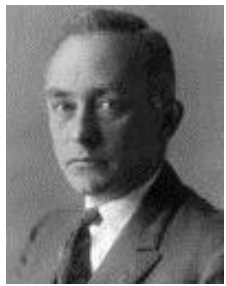

**Max Born**

In 1926 Max Born, studying the scattering of a particle performed by a given potential, suggested the probabilistic interpretation of the eigenfunction [11, 23-25]. Precisely, the squared modulus of the quantum function $|\psi|^2$, which is always real, represents the probability density to find the particle in a given part of the space (we suppose $|\psi|^2$ time-independent). This interpretation was proved by more experimental results and with the Solvay Congress in 1927 was accepted by the scientific community. The condition that the function $|\psi|^2$ has to satisfy to be a probability density is:

$$\int_{-\infty}^{\infty} \rho(\boldsymbol{r})d\boldsymbol{r} = 1 \qquad (\boldsymbol{9.a})$$

or the probability to find the particle in the whole space must be equals to one. This condition implies that all the eigenfunctions of the hydrogen atom have to be normalized according to:

$$\int_{-\infty}^{\infty} \psi^*(\boldsymbol{r})\psi(\boldsymbol{r})d\boldsymbol{r} = 1 \qquad (\boldsymbol{9.b})$$

The **(9.b)** represents the boundary condition for each Schrödinger problem; we conclude that the eigenfunctions are an orthonormal base for the Hamiltonian operator and they are the elements of the Hilbert space $L^2(\mathbb{R}, \mathbb{C})$ we met in the paragraph 1.8 (this symbol means continues functions with complex values, derivable at list two times and squared integrable).

Go back to the radial functions $R_{n,l}(\boldsymbol{r})$; we note that the orbitals $ns$ have $(n-1)$ nodes where the function is zero. That means the density function $|ns|^2$ has the same number of nodes and around the atomic core there are $(n-1)$ points where the probability to find the particle is zero.

The same considerations may be done for all others orbitals of hydrogen atom. Another relevant result arising from the Schrodinger theory is that increasing the main quantum number the eigenfunctions tend to zero more slowly (as already said in section 6). In other words, the increase of the main quantum number leads to a deeper penetration of the potential barrier due to the atomic core (that in the case of the hydrogen atom has an hyperbolic shape). This phenomenon is justified by the fact that increasing the main quantum number the electron energy increases becoming so *able* to penetrate more and more the potential barrier.

The probabilistic interpretation of the eigenfunction is simple and solves in a definitive manner the doubt that Schrödinger had about the physical interpretation of the wave function for the free particle: the wave packet is not the deterioration over the time of its physical reality but, rather, is simply the loss of information on its position.

<p style="text-align:center">*****</p>

In the following we will use the Dirac notation on the bra and ket:

- $|\psi(r,t)\rangle$ is a ket vector representing the state of quantum particle. It is the eigenfunction solution of the quantum equation of motion.

- $\langle\psi(r,t)|$ is a bra vector. It is the adjoint of the ket vector.

- $\langle \psi_1(r,t) | \psi_2(r,t) \rangle = \int_{-\infty}^{\infty} \psi_1(r,t)^\dagger \psi_2(r,t) dr dt$

- $\langle \psi_1(r,t) | H | \psi_2(r,t) \rangle = \int_{-\infty}^{\infty} \psi_1(r,t)^\dagger H \psi_2(r,t) dr dt$, where H is a quantum operator.

## 10. The Heisenberg Equations

In the Schrödinger picture the state of a quantum system is a vector satisfying the equation:

$$i\hbar \frac{\partial}{\partial t} |\psi(r,t)\rangle = H(t)|\psi(r,t)\rangle$$

Let's introduce the time-evolution operator $U$ that transform the state $|\psi(r,t_0)\rangle$ at time $t_0$ to the state $|\psi(r,t)\rangle$ at the time $t$:

$$|\psi(r,t)\rangle = U(t,t_0)|\psi(r,t_0)\rangle \qquad (\mathbf{10.a})$$

This operator satisfies the properties:

a) $U(t_0,t_0) = \mathbb{1}$

b) $U(t,t_0)[c_1|\psi(r,t_0)\rangle + c_2|\psi(r,t_0)\rangle] = c_1|\psi(r,t)\rangle + c_2|\psi(r,t)\rangle$

c) $U(t,t_0) = U(t,t_1)U(t_1,t_0)$

From the property c) it follows that:

$$U(t_0,t_0) = U(t_0,t)U(t,t_0) = \mathbb{1}$$

or:

$$U(t_0,t) = U^{-1}(t,t_0) \quad \Rightarrow \quad U(t_0,t) = U^\dagger(t_0,t)$$

which states that the time-evolution operator is unitary. To prove this property we consider two states $|\psi_1(r,t)\rangle$ and $|\psi_2(r,t)\rangle$; applying the time-dependent Schrödinger equation we get:

$$\frac{\partial}{\partial t}\langle \psi_1(r,t) | \psi_2(r,t) \rangle = \langle \dot{\psi}_1(r,t) | \psi_2(r,t) \rangle + \langle \psi_1(r,t) | \dot{\psi}_2(r,t) \rangle =$$

$$= \frac{i}{\hbar}[-\langle \psi_1(r,t)|H|\psi_2(r,t)\rangle + \langle \psi_1(r,t)|H|\psi_2(r,t)\rangle] = 0$$

This result is based on the fact that the matrix elements $\langle \psi_1(r,t)|\psi_2(r,t)\rangle$ are time-independent. So, we can state that:

$$U^\dagger(t_0,t) = U^{-1}(t,t_0)$$

Le's replace now the expression of eigenspace $|\psi(r,t)\rangle$ given by the $(\mathbf{10.a})$ in the time-dependent Schrödinger equation:

$$i\hbar \frac{\partial}{\partial t}|\psi(r,t)\rangle = i\hbar \frac{\partial}{\partial t}[U(t,t_0)|\psi(r,t_0)\rangle] = HU(t,t_0)|\psi(r,t_0)\rangle = H|\psi(r,t_0)\rangle$$

Taking into account the two central members of this equation we can write:

$$i\hbar \frac{\partial}{\partial t}U(t,t_0) = HU(t,t_0) \qquad (\mathbf{10.b})$$

If the Hamiltonian operator does not explicitly depend on time (steady states) then the eigenvectors get the form:

$$|\psi(r,t)\rangle = |\psi(r)\rangle exp\left\{-i\frac{E(t-t_0)}{\hbar}\right\} \qquad (\mathbf{10.c})$$

As previously proved the exponential term has unitary norm and does not affect the eigenvalue calculation. Moreover, being a complex exponential it has periodicity of $2\pi$ and its action on the vector $|\psi(r)\rangle$ is that of a rotation in a given direction with a frequency $\nu = E/\hbar$:

$$f(t) = exp\{-2\pi i\nu\}(t-t_0)$$

Therefore, in the Schrödinger picture if the Hamiltonian does not explicitly depend on time then the vectors $|\psi(r)\rangle$ evolve according an harmonic function $f(t)$, which does not modify their length but change their direction in the space. In others words, the Schrödinger eigenstates of a steady state rotate in the Hilbert space in a given direction with a frequency of $\nu = E/\hbar$. Greater is the energy of the steady state and greater will be its angular rotation speed. Concluding, in the Schrödinger picture the operators associated to the observables are fix over the time while their eigenvectors, representing the steady states, are evolving. We can so well understand as the Schrödinger picture is *far* from the laws of the classical mechanics, where are the physical observables to evolve over the time.

Taking once in consideration the (**10.c**), the exponential term can be also rewritten as:

$$|\psi(r,t)\rangle = exp\left\{-i\frac{H(t-t_0)}{\hbar}\right\}|\psi(r,t_0)\rangle$$

Comparing this equation with (**10.a**) is possible to give an explicit form of the time-evolution operator $U(t,t_0)$:

$$U(t,t_0) = exp\left\{-i\frac{H(t-t_0)}{\hbar}\right\} \qquad (\mathbf{10.d})$$

Performing the time-derivative of this expression we get:

$$\frac{\partial}{\partial t}U(t,t_0) = \frac{\partial}{\partial t}exp\left\{-i\frac{H(t-t_0)}{\hbar}\right\} = -\frac{i}{\hbar}Hexp\left\{-i\frac{H(t-t_0)}{\hbar}\right\} = -\frac{i}{\hbar}HU(t,t_0)$$

that rearranged gives just the equation (**10.b**), confirming that the (**10.d**) is well set.

If the Hamiltonian operator does not explicitly depend on time it can be developed in Taylor series:

Performing the development we get:

$$U(t,t_0) = \sum_{k=0}^{\infty}\frac{\left[-i\frac{H(t-t_0)}{\hbar}\right]^k}{k!}$$

Cutting the series at $k = 2$ it's obtained:

$$U(t, t_0) = 1 - i\frac{H(t - t_0)}{\hbar} - \frac{1}{2}\left[-i\frac{H(t - t_0)}{\hbar}\right]^2$$

Applying this operator to the ket $|\psi(r, t_0)\rangle$:

$$U(t, t_0)|\psi(r, t_0)\rangle = |\psi(r, t_0)\rangle - \frac{iE}{\hbar}(t - t_0)|\psi(r, t_0)\rangle - \frac{1}{2}\frac{E^2}{\hbar^2}(t - t_0)^2|\psi(r, t_0)\rangle$$

This example shows how the operator $U$ given by the **(10.d)** acts on the ket. Recalling the form of the classical mechanics laws, Heisenberg formulated a new picture of quantum mechanics, usually known as matrix mechanics, where the operators evolve over the time and the vectors remain stationaries. It should be clear that Heisenberg formulated its theory without knowing the Schrödinger work; the two theories were formulated by the two physicists almost simultaneously (1925-1926) [26].

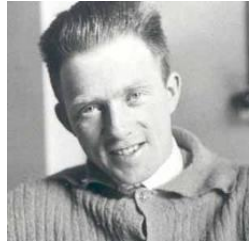

**Werner Heisenberg**

The Heisenberg picture is very similar to that of the classical mechanics where the observables evolve over the time. As a fact the Heisenberg picture follows an approach diametrically opposed than that used by Schrödinger even if, as will be proved further, they get the same eigenvalues. This difference arise from the two distinct starting points: Schrödinger used the De Broglie hypothesis (of wave nature) while Heisenberg developed its theory starting from the uncertainty principle which is based on the non-commutative algebra [22].

Let's consider an unitary operator that brings back to the initial state the vectors $|\psi(r, t)\rangle$; this operator is just the adjoint of the operator $U(t, t_0)$ previously introduced:

$$U^\dagger(t, t_0) = U(t_0, t)$$

From now on we denote by s and H the Schrödinger and Heisenberg pictures. Starting from the last equality we write the following equations:

$$|\psi_s(r, t)\rangle = U(t, t_0)|\psi_H(r, t)\rangle$$

$$|\psi_H(r, t)\rangle = U^\dagger(t_0, t)|\psi_s(r, t)\rangle$$

The two pictures must be equivalent and they must lead to the same expectation mean values:

$$\langle\psi_s|A_s|\psi_s\rangle = \langle\psi_H|A_H|\psi_H\rangle = \langle\psi_H|U^\dagger A_s U|\psi_H\rangle$$

where $A$ is an operator associated to the observable $a$. From the equality of the expectation mean values we get:

$$A_H = U^\dagger A_s U = U^{-1} A_s U \qquad (\textbf{10.e})$$

because $U$ is unitary and so $U^\dagger = U^{-1}$. If the operator is represented by a matrix then the relation (**10.e**) is nothing more than a similarity relation between matrices. The (**10.e**) shows that in the Heisenberg picture the operator $A$ depends always over the time since $U$ is a time-operator. The operators are time-dependent and we have to find the equation describing this dependence. To do this we perform the time derivative of (**10.e**):

$$\frac{dA_H}{dt} = \left(\frac{dU^\dagger}{dt}\right) A_s U + U^\dagger \left(\frac{dA_s}{dt}\right) U + U^\dagger A_s \left(\frac{dU}{dt}\right)$$

and using the equation (**10.b**) we get:

$$\frac{dA_H}{dt} = \frac{i}{\hbar} H_s U^\dagger A_s U + U^\dagger \left(\frac{dA_s}{dt}\right) U - \frac{i}{\hbar} U^\dagger A_s H_s U = \frac{i}{\hbar} U^\dagger H_s A_s U + U^\dagger \left(\frac{dA_s}{dt}\right) U - \frac{i}{\hbar} U^\dagger A_s H_s U$$

where some steps are justified by the fact that $H_s$ is self-adjoint. Using the inverse relationship of (**10.e**) we obtain:

$$H_s = U H_H U^\dagger \quad and \quad A_s = U A_H U^\dagger$$

Replacing these results in the last expression of $\frac{dA_H}{dt}$ we get:

$$\frac{dA_H}{dt} = \frac{i}{\hbar} U^\dagger U A_H U^\dagger A_s U + U^\dagger \left(\frac{dA_s}{dt}\right) U - \frac{i}{\hbar} U^\dagger A_s U H_H U^\dagger U =$$

$$= \frac{i}{\hbar} H_H A_H + U^\dagger \left(\frac{dA_s}{dt}\right) U - \frac{i}{\hbar} A_H H_H =$$

$$= \frac{i}{\hbar} [H_H, A_H] + U^\dagger \left(\frac{dA_s}{dt}\right) U$$

where $[H_H, A_H]$ is the commutator of the two operators $H_H$ and $A_H$. Reworking the equation obtained and recalling that $[H_H, A_H] = -[A_H, H_H,]$ we get:

$$i\hbar \frac{dA_H}{dt} = [A_H, H_H] + i\hbar \left(\frac{dA_H}{dt}\right) \qquad (\textbf{10.f})$$

where we keep in mind that:

$$U^\dagger \left(\frac{dA_s}{dt}\right) U = \left(\frac{dA_H}{dt}\right)$$

The (**10.f**) is the equation of motion for the quantum operator in the Heisenberg picture. This equation is equivalent to the time-dependent Schrödinger one! If the operator $A_s$ does not explicitly depend on the time then also the derivative $\frac{dA_H}{dt}$ is zero and the equation becomes:

$$i\hbar \frac{dA_H}{dt} = [A_H, H_H] \qquad (\textbf{10.g})$$

If the operator $A$ in the Heisenberg picture commutes with the Hamiltonian $H_H$ then $\frac{dA_H}{dt} = 0$. That means the operator $A$ is a constant of the motion and its expectation mean value will remain constant over the time.

One of the most important evidence contained in equation (**10.f**) is the presence of a commutator: in the Heisenberg picture the operators obey to the non-commutative algebra. This result is true also for the operators in the Schrödinger picture but in a non-explicit form!

To arrive at the final formulation of the Heisenberg equation we have to consider the properties of the commutators. In this regard let be $A$, $B$ and $C$ three operators; then the following relations are hold:

a) $[A, B] = -[B, A]$

b) $[A, B + C] = [A, B] + [A, C]$

c) $[A, BC] = [A, B]C + B[A, C]$

d) $\big[A, [B, C]\big] + \big[B, [C, A]\big] + \big[C, [A, B]\big] = 0$

e) $[A, B]^\dagger = [B^\dagger, A^\dagger]$

In section 5 we proved that to the classical momentum $p$ is associated the operator $\hat{p} = i\hbar\nabla$. Without affecting any generality is possible to associate also the operator $\hat{p} = -i\hbar\nabla$. In fact this operator is always hermitian [21]:

$$\hat{p}^\dagger = i\hbar\nabla^\dagger = i\hbar(-\nabla) = -i\hbar\nabla = \hat{p}$$

Moreover, the eigenvalue equations $\left(\frac{\hat{p}^2}{2m} + U\right)\psi = E\psi$ and $\hat{p}\psi = p\psi$ do not change using the operator $\hat{p} = -i\hbar\nabla$. What is changed is the sign of the commutator. But this change does not alter the physical meaning of the result being the commutator an abstract entity that cannot be associated to physical observables. Recalling the main commutators introduced in section 8 for the impulse operator $\hat{p} = -i\hbar\nabla$, we have:

$$[q, p] = i\hbar$$
$$[q_i, q_i] = 0$$
$$[p_i, p_i] = 0$$

where $q$ and $p$ are position and impulse coordinates. Let's suppose to neglect the difference between the non-commutative algebra of quantum operators and the commutative algebra of classical observables. Is so possible writing the commutation relations as:

a) $\quad [q_i, p_i] = i\hbar\frac{\partial p_i}{\partial p_i}$

b)      $[p_i, q_i] = -i\hbar \frac{\partial q_i}{\partial q_i}$

c)    $[q_i, q_i] = i\hbar \frac{\partial q_i}{\partial p_i} = 0$

d)    $[p_i, p_i] = i\hbar \frac{\partial p_i}{\partial q_i} = 0$

Moreover, if $F$ is a generic function of conjugated coordinates $p$ and $q$ we have:

$$\begin{cases} [q_i, F] = i\hbar \frac{\partial F}{\partial p_i} \\ [p_i, F] = -i\hbar \frac{\partial F}{\partial q_i} \end{cases} \qquad \textbf{(10.h)}$$

obtained by analogy with relations c) and b). The calculation we are using is based on the Poisson bracket of the classical mechanics:

$$\{u, v\} = \left[ \frac{\partial u}{\partial q_i} \frac{\partial v}{\partial p_i} - \frac{\partial u}{\partial p_i} \frac{\partial v}{\partial q_i} \right]$$

with $u$ and $v$ whatever classical variables. The commutators **(10.h)** tends to the classical ones when $\hbar \to 0$; the Heisenberg equations of motion are based on the Bohr correspondence principle.

Using equation **(10.g)** is possible to calculate the equations of time-evolution for the operators $q_i$ and $p_i$:

$$\frac{dq_i}{dt} = \frac{1}{i\hbar} [q_i, H] = \frac{1}{i\hbar} i\hbar \frac{\partial H}{\partial p_i} = \frac{\partial H}{\partial p_i}$$

$$\frac{dp_i}{dt} = \frac{1}{i\hbar} [p_i, H] = \frac{1}{i\hbar} \left( -i\hbar \frac{\partial H}{\partial q_i} \right) = -\frac{\partial H}{\partial q_i}$$

where $H$ is a function of the two conjugated variables $q_i$ and $p_i$. Summarizing:

$$\begin{cases} \dfrac{dq_i}{dt} = \dfrac{\partial H}{\partial p_i} \\ \dfrac{dp_i}{dt} = -\dfrac{\partial H}{\partial q_i} \end{cases} \qquad \textbf{(10.i)}$$

These are the Heisenberg equations and their form is analogous to that of Hamilton of classical mechanics with the analogy:

$$\{A, H\} \quad \Rightarrow \quad \frac{1}{i\hbar}[A, H]$$

The formalism of the Poisson bracket corresponds to the commutator of the hermitian operators. We have to recall that the quantities in equations **(10.i)** are matrices and the operator $H$ is a function of all the matrices associated to the conjugated observables:

$$H = H(q_1, \dots, q_n, p_1, \dots, p_n)$$

So, in the Heisenberg picture the connection with the classical mechanics is well clear. Equations **(10.i)** have been obtained under the hypothesis that the operators $q_i$ and $p_i$ are not explicitly time-dependent; otherwise, in **(10.i)** we need add the terms $\frac{dq_i}{dt}$ and $\frac{dp_i}{dt}$ according to equation **(10.f)**.

Let's return to the equation **(10.e)**:

$$A_H = U^\dagger A_s U = U^{-1} A_s U$$

Supposing that the operator $A_s$ in the Schrödinger picture is not explicitly time-dependent we can state that in the Heisenberg picture the same operator depends over the time. Moreover, using the explicit form of $U$ given by the **(10.d)**, the operator rotates in the opposite direction respect that of the eigenvectors in the Schrödinger picture. Also this behavior, in the Heisenberg picture, is perfectly analogous to that of the observables of the classical mechanics!

Heisenberg developed his theory starting from very different assumptions! Inspired by the Bohr theory, Heisenberg introduced a new steering hypothesis according which the trajectory of the electron in the hydrogen atom cannot be known with arbitrary precision but the second low of the dynamic is however preserved. The new theory must be able to explain the quantum transitions representing the hydrogen spectrum. If we denote by $r(n, t)$ the classical equation of the stationary orbits provided by the Bohr theory then its explicit form can be represented by a Fourier series:

$$r(n, t) = \sum_{\alpha=-\infty}^{\infty} c_\alpha(n) exp\{i\alpha\omega(n)t\}$$

where $\omega(n)$ is the angular velocity of electron in the n[th] state. Heisenberg proposed, on the basis of the correspondence principle, to connect each component of the Fourier series with a given electron transition. That means to replace the classical terms $c_\alpha(n)exp\{i\alpha\omega(n)t\}$ with another of the type $c_\alpha(n,m)exp\{i\alpha\omega(n,m)t\}$, where $m$ labels a steady state different from the n[th] one, so that every electronic transition from the state $n$ to the state $m$ is contemplated. In this way the classical function $r(n,t)$ is replaced by a new mathematical entity whose structure is identical to that of a matrix with components $r_{nm}$. If the electronic transition is allowed then matrix component will be different from zero. Heisenberg arrived to the concept of operator associated to a classical observable! A fundamental aspect emerges from the Heisenberg theory: the electron position is known when a transition between two states occurs! That is just what we called collapse of a quantum state during a measurement.

The electron impulse can be easily calculated differentiating the elements of the matrix $r_{nm}$:

$$p_{nm} = m\dot{r}_{nm} = mc_\alpha(n,m)i\alpha\omega(n,m)exp\{i\alpha\omega(n,m)t\} = im\omega_{nm}r_{nm}$$

On the basis of these results Heisenberg also reworked the Bohr-Sommerfeld quantization rule:

$$\oint p dq = 2\pi n\hbar$$

Since the electronic transitions are discrete, Heisenberg made the assumption that the rule must be rewritten as difference between integrals concerning two neighboring states:

$$\oint p dq\Big|_n - \oint p dq\Big|_{n-1} = 2\pi\hbar$$

Replacing the expression previously found for the conjugated variables we get:

$$2m\sum_{\alpha=0}^{\infty}\{|c_\alpha(n+\alpha)|^2\omega(n+\alpha,n) - |c_\alpha(n-\alpha)|^2\omega(n-\alpha,n)\} = \hbar$$

This relation gives the quantum rule according which the amplitudes of the spectra lines are related among them.

What about the matrix $\omega_{nm}$? We know that their elements are connected to all electronic transitions; we know also that the term $\hbar\omega$ is the energy. Therefore, we can write the energy value related to a transition from the state $m$ to the state $n$ as:

$$\hbar\omega_{nm} = E_n - E_m$$

We obtained a new matrix $H_{nm}$ representing the transition energies of the hydrogen atom whose diagonal elements are all zero. That means when the electron is in a possible steady state, its energy is constant; this suggest that the steady states of the atomic electron may be represented by a diagonal matrix $H_{nn}$ by which the last relation can be rewritten as:

$$\hbar\omega_{nm} = H_{nn} - H_{mm}$$

The rest of the Heisenberg theory is developed exactly on the basis of the Hamiltonian mechanics (that is the assumption to preserve the validity of the second law of the dynamic made by Heisenberg). For instance, if we replace in the first classical Hamiltonian equation the Heisenberg operator $q_{mn}$ written above we get:

$$\dot{q}_{nm} = \frac{i}{\hbar}[(E_n - E_m)c_{nm}exp\{i\omega_{mn}t\}] = \frac{i}{\hbar}(H_{nn}q_{nm} - q_{nm}H_{mm})$$

which is equivalent to the well-known matrix equation:

$$\frac{d\hat{q}}{dt} = \frac{i}{\hbar}[H, q]$$

## 11. The Heisenberg-Pauli Factorization Method

The resolution of a quantum equation in the picture of Heisenberg mechanics requires the use of the factorization method which consists in writing the hermitian operator as product of two terms, one the adjoint of the other, (not necessarily Hermitian):

$$\hat{A} = \hat{a}^\dagger\hat{a} + \lambda\mathbb{1} \qquad (11.a)$$

where $\lambda$ is the eigenvalue of the operator $\hat{A}$. In the case $\hat{A}$ admits more than a factorization must be chosen the one that gives the highest eigenvalue [27]. The operators $\hat{a}$ and $\hat{a}^\dagger$ are called creator and annihilation operators. Unfortunately does not exist an algorithm to find such a factorization, but the success of calculation depend on the ability of who is facing the problem to be studied!

Although the goal of the Heisenberg theory is that to calculate of the eigenvalues of the Hermitian operator, using the factorization method we can find also the explicit form of the ket.

## 12. The Hydrogen Atom in the Picture of Matrix Mechanics

Let's study now the hydrogen atom in the Heisenberg picture. Since the Heisenberg theory is based on the algebra of hermitian matrices we are interested in the calculation of the energy eigenvalues!

The Hamiltonian operator of the atom is:

$$H = \frac{p^2}{2m} + \frac{L^2}{2mr^2} - \frac{e^2}{r}$$

where $L$ is the angular momentum operator of the electron, while the potential energy has been written omitting the constant term $4\pi\varepsilon_0$. Recalling the result obtained in section 8 for the squared angular momentum operator, the Hamiltonian can be reworked as:

$$H = \frac{p^2}{2m} + \frac{\hbar^2 l(l+1)}{2mr^2} - \frac{e^2}{r} \qquad (\boldsymbol{12.a})$$

We apply now the factorization method with the aim to write the operator $H$ as product of complex conjugated matrices:

$$H_n = \hat{a}^\dagger{}_n \hat{a}_n + \lambda_n \mathbb{1}$$

which concerns the generic $n^{th}$ state of hydrogen atom. We set the creator and annihilation operators as:

$$\hat{a}_n = \frac{1}{\sqrt{2}}\left[ p + i\left(\alpha_n + \beta_n \frac{1}{r}\right)\right]$$

$$\hat{a}^\dagger{}_n = \frac{1}{\sqrt{2}}\left[ p + i\left(\alpha_n - \beta_n \frac{1}{r}\right)\right]$$

where $\alpha_n$ and $\beta_n$ are real numbers that have to be calculated. Performing the operator product $\hat{a}^\dagger{}_n \hat{a}_n$ we get:

$$\hat{a}^\dagger{}_n \hat{a}_n = \frac{1}{2m}\left[ p^2 + \alpha_n^2 + \frac{2\alpha_n\beta_n}{r} + \frac{(\beta_n^2 - \hbar\beta_n)}{r^2}\right] \qquad (\boldsymbol{12.b})$$

and the factorization becomes:

$$H_n = \hat{a}^\dagger{}_n \hat{a}_n + \lambda_n \mathbb{1} = \frac{1}{2m}\left[ p^2 + \alpha_n^2 + \frac{2\alpha_n\beta_n}{r} + \frac{(\beta_n^2 - \hbar\beta_n)}{r^2}\right] + \lambda_n \mathbb{1} \qquad (\boldsymbol{12.c})$$

Comparing (**12.c**) and (**12.a**):

$$\frac{2\alpha_n\beta_n}{r} = -\frac{2me^2}{r} \quad , \quad \frac{(\beta_n^2 - \hbar\beta_n)}{2mr^2} = \frac{\hbar^2 l(l+1)}{2mr^2} \quad , \quad \lambda_n = -\frac{\alpha_n^2}{2m}$$

We have found the numerical constants $\alpha_n$ and $\beta_n$ using our ability with the algebraic calculus. For the fundamental state these numbers could be the following:

$$\alpha_0 = \frac{me^2}{\hbar l} \quad , \quad \beta_0 = -\hbar l$$

which gives the eigenvalue $E_0$:

$$E_0 = -\frac{\alpha_0^2}{2m} = -\frac{me^4}{2\hbar^2 l^2}$$

But for the fundamental state we know that $l = 0$ and so the obtained result is meaningless! We can try then with a new set of numbers:

$$\alpha_0 = -\frac{me^2}{\hbar(l+1)} \quad , \quad \beta_0 = \hbar(l+1)$$

which gives the new eigenvalue $E_0$:

$$E_0 = -\frac{\alpha_0^2}{2m} = -\frac{me^4}{2\hbar^2(l+1)^2}$$

This value is the same of that found by Bohr and Schrödinger. Proceeding according the iterative calculus of the factorization method we get for the n[th] state the following set of numbers:

$$\alpha_n = -\frac{me^2}{\hbar(l+n)} \quad , \quad \beta_0 = \hbar(l+n)$$

to which corresponds the energy:

$$E_{l,n} = -\frac{\alpha_n^2}{2m} = -\frac{me^4}{2\hbar^2(l+n)^2}$$

$n$ is a positive integer and $l$ must be a non-negative integer [27].

## 13. The Equivalence of Wave and Matrix Mechanics

The mathematical equivalence between the two formalisms was studied and solved by Schrödinger in 1926 [15, 21]. The problem of matrix mechanics is that to find hermitian matrices $P$ and $Q$, associated to the linear momentum $p$ and position $q$, such as to satisfy the commutator $[P.Q] = i\hbar$ and such that the matrix $H$, function of $P$ and $Q$, is diagonal. In the case they had not this form will be always possible to find an invertible matrix $S$ such that $S^{-1}HS$ is diagonal. If $X$ is a vector of the eigenspace associated to the matrix $H$ respect to a given base (from the algebra we know that a base of the eigenspace of $H$ is the set of the column vector forming the matrix $S$), then is satisfied the following eigenvalues equation:

$$\sum_{n'} H_{nn'} X_{n'} = \lambda X_n \quad n = 1, 2, \dots \qquad (13.a)$$

where $H_{nn'}$ are the components of the matrix $H$, while $X_{n'}$ are those of the column vector $X'$. In this context the Heisenberg theory is clear and well defined. In the wave mechanics, instead, the basic problem is represented by the following differential equation:

$$H\psi(q_1, \dots, q_k) = \lambda \psi(q_1, \dots, q_k) \qquad (13.b)$$

whose structure is similar to the **(13.a)**. However, the **(13.a)** is a linear algebraic equation while the second one must be a differential equation, even if both are eigenvalues equations. We want now to find their mathematical equivalence, if it exists! We begin by noting that the index $n$ of the (**13.a**) is completely analogous to the index $k$ of the (**13.b**), which is the dimension of the configuration space $\boldsymbol{\Omega}$ associated to the differential operator $H$. We can therefore assuming that the summation $\sum_n$ is analogous to the volume integral calculated over the space $\boldsymbol{\Omega}$:

$$\sum_{n'} \ldots \quad \leftrightarrow \quad \int_\Omega dq'_1 \cdots dq'_k = \int_\Omega dV$$

From this relationship it follows also that:

$$X_n \quad \leftrightarrow \quad H_{nn'} X_{n'}$$

$$\psi(q_1, \ldots, q_k) \quad \leftrightarrow \quad \int_\Omega H(q_1, \ldots, q_k; \; q'_1, \ldots, q'_k) \, dq'_1 \cdots dq'_k$$

We can then rewrite the eigenvalues problem (**13.a**) as:

$$\int_\Omega H(q_1, \ldots, q_k; \; q'_1, \ldots, q'_k) \, \psi(q'_1, \ldots, 'q_k) dq'_1 \cdots dq'_k = \lambda\psi(q_1, \ldots, q_k) \qquad (\mathbf{13.c})$$

Therefore, starting from an algebraic equation we obtained an integral equation that, at least in the first instance, has nothing in common with the differential equation (**13.b**). However, in the field of the generalized functions, to which belongs that of Dirac, is always possible representing a differential operator as an integral:

$$\frac{\partial^n}{\partial q^n}\psi(q) = \frac{\partial^n}{\partial q^n}\int_{-\infty}^{\infty}\delta(q-q')\psi(q')dq' = \int_{-\infty}^{\infty}\frac{\partial^n}{\partial q^n}\delta(q-q')\psi(q')dq' =$$

$$= \int_{-\infty}^{\infty}\frac{\partial^n}{\partial q^n}\delta^{(n)}(q-q')\psi(q')dq'$$

The derivative has been brought under the integral sign because it does not act on the integration variable. The functional $\delta^{(n)}(q-q')$ is defined Kernel of the integral operator. Using this representation we can rewrite equation (**13.c**) as:

$$\int_\Omega H(q_1, \ldots, q_k; \; q'_1, \ldots, q'_k) \, \psi(q'_1, \ldots, 'q_k) dq'_1 \cdots dq'_k =$$

$$= \int_{-\infty}^{\infty}\cdots\int_{-\infty}^{\infty}\delta(q_1-q'_1)\cdots\delta(q_k-q'_k)\,\psi(q'_1, \ldots, 'q_k)dq'_1 \cdots dq'_k$$

$$= \lambda\psi(q_1, \ldots, q_k) \qquad (\mathbf{7.9.d})$$

We conclude that to the vector $X_n$ corresponds the vectorial function $\psi$, to the summation $\sum_{n'}$ corresponds the integral $\int_\Omega dq'_1 \cdots dq'_k$, to the summation index $n'$ correspond the coordinates $q'_1, \ldots, q'_k$, to the index $n$ correspond the coordinates $q_1, \ldots, q_k$ and to the matrix elements $H_{nn'}$ corresponds the Kernel $\delta(q_1-q'_1)\cdots\delta(q_k-q'_k)$. The mathematical equivalence of the two theories is so proved.

<p style="text-align:center">*　*　*</p>

Despite the elegance of the formalism and its close connection with the Hamiltonian mechanics, the matrix mechanics has never been used systematically by the physicists, as it has been for the Schrödinger formalism. This is due to the fact that the physicists usually prefer solving problems using differential equations and to the strong development that the analysis has had in the XX century. The Heisenberg theory renounces to localize the electron in the space and time and focuses the attention on measurable quantities. Is not a case that a brilliant mind like Pauli had to perform an intense work to solve by matrix mechanics the problem of the hydrogen atom [27].

## 14. The Quantum Relativistic Theory

The quantum theory developed so far is based on a non-relativistic approach since the velocities of the electrons, at least for light atoms, are lower than the speed of light. For instance, considering the Bohr model of the hydrogen atom the velocity of the electron is given by:

$$v = \frac{e^2}{4\pi\varepsilon_0 n\hbar}$$

Supposing the electron in the fundamental state we get a velocity of $2.18 \ 10^6 m/s$ or about 1/100 of that of light! Is not a coincidence that the non-relativistic quantum theory leads to results in good agreement with the experimental ones. It is then natural to wonder why study the atomic theory on the basis of relativity, with the risk of further complicating the mathematical formalism, when the available models already give us good results. We will see the relativistic approach leads to the formulation of equations characterized by an elegant and powerful formalism, able to predict the property of spin and the physical phenomenon arising from it [28]. For instance, the formulation of a relativistic equation for the hydrogen atom electron allows calculating directly the spin-orbit coupling operator $\hat{J}^2$, without to have to determine it *ad hoc* as must be done in the Schrodinger theory. In other words, we can state that relativistic equations lead to theoretical results fully in agreement with the experimental ones, without the need to introduce any further hypothesis. To this we must add that the introduction of the relativity in the quantum mechanics allowed the physicists of the first half of XX century to open a new field of science, known as quantum field theory.

Historically, the first successful application of relativity theory to the quantum mechanics is due to P.A.M. Dirac that, in 1928, published the article that introduced its famous equation for the free electron and for the hydrogen atom. This last allowed to explain correctly the fine structure of the spectrum of hydrogen atom and to predict the existence of anti-particles, whose discovery would take place a few years later. The most interesting aspect of the Dirac equation is that it was formulated following an approach mainly mathematic. Often, the *tools* of the algebra or those of the analysis are abstract entities, create irrespective of the fact that may or may not have practical applications! Dirac remarked how *often the Nature is described and let it explain just by the more abstract equations and as such the most elegant.*

## 14.1. The Dirac Equation

The physical-mathematical discussion that follows in this section is freely obtained from the work of Dirac on the formulation of the relativistic electron equation.

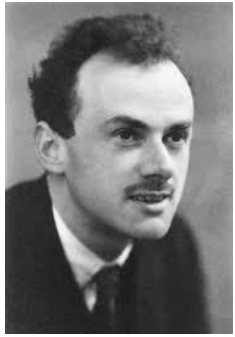

**P.A.M Dirac**

We are seeking a relativistic equation where all the coordinates are symmetric. Moreover, the equation must be of first order so as to avoid getting a negative probability density, that is what occurs using the relativistic Klein-Gordon equation [29]. If we need a first time-derivative it follows that also the derivative performed respect the position coordinates must be of the first order. This is the starting point to get the Dirac equation, supported by an exclusively mathematics requirement arising from the condition of invariance versus the Lorentz transformations. For a free electron the equation can be written as:

$$\sum_{k=0}^{3} \alpha_k \frac{\partial}{\partial x^k} |\psi\rangle = 0 \qquad (\mathbf{14.1.a})$$

The coordinate $x^0$ is given by $ct$ and thus the first term of the summation can be rewritten as $\alpha_0 \frac{\partial}{\partial ct}$; setting $\alpha_0 = 1$ the (**14.1a**) becomes:

$$\left( \frac{1}{c} \frac{\partial}{\partial t} + \sum_{k=1}^{3} \alpha_k \frac{\partial}{\partial x^k} \right) |\psi\rangle = 0 \qquad (\mathbf{14.1.b})$$

We observe that all the elements of the summation have the dimension of an inverse length and are proportional to the operatorial components of the four-vector $\hat{p}$:

$$\hat{p} = -i\hbar \left( \frac{1}{c} \frac{\partial}{\partial t}, \frac{\partial}{\partial x^1}, \frac{\partial}{\partial x^2}, \frac{\partial}{\partial x^3} \right)$$

The coefficients $\alpha_k$ are numbers and do not depend by the space-time coordinates nor by the impulse. These coefficients, therefore, commute both with the coordinates $x^k$ and the impulse $\hat{p}$.

In principle the equation **(14.1.b)** may contains also a numerical constant having the dimension of an inverse length. We can choose this constant as proportional to the term $\frac{mc}{\hbar}$ whose dimension is just that of an inverse length. Since the relativistic equation we are seeking can be complex, we decide to write the constant term as $\frac{imc}{\hbar}\beta$, where $\beta$ is a numerical coefficient. The **(14.1.b)** may be rewritten in a complete form as:

$$\left(\frac{1}{c}\frac{\partial}{\partial t} + \sum_{k=1}^{3}\alpha_k\frac{\partial}{\partial x^k} + \frac{imc}{\hbar}\beta\right)|\psi\rangle = 0 \qquad (\mathbf{14.1.c})$$

Because $\beta$ is a numerical coefficient it must commute with $x^k$ and $\hat{p}$. The numerical coefficients $\alpha_k$ and $\beta$ seem describing a new degree of freedom of the electron that in the non-relativistic discussion does not arise. Multiplying both members of the **(14.1.c)** times $\frac{\hbar}{i} = -i\hbar$ we get:

$$\left(-\frac{1}{c}i\hbar\frac{\partial}{\partial t} - \sum_{k=1}^{3}\alpha_k i\hbar\frac{\partial}{\partial x^k} + \beta mc\right)|\psi\rangle = 0 \qquad (\mathbf{14.1.d})$$

It's evident that the first term of the summation recall the four-vector impulse $\hat{p}$:

$$(p_0 + \alpha_1 p_1 + \alpha_2 p_2 + \alpha_3 p_3 + \beta mc)|\psi\rangle = 0 \qquad (\mathbf{14.1.d'})$$

Dirac proposed to consider the eigenfunction $|\psi\rangle$ as a vector with more components; under this hypothesis the coefficients $\alpha_k$ and $\beta$ are square matrices having the same dimension of the vector $|\psi\rangle$. The equation **(14.1.d)** can be also written as:

$$i\hbar\frac{\partial}{\partial t}|\psi\rangle = \left(-\sum_{k=1}^{3}c\alpha_k i\hbar\frac{\partial}{\partial x^k} + \beta mc^2\right)|\psi\rangle \qquad (\mathbf{14.1.e})$$

that resembles the time-dependent Schrödinger equation where the Hamiltonian operator is:

$$H = -\sum_{k=1}^{3}c\alpha_k i\hbar\frac{\partial}{\partial x^k} + \beta mc^2 \qquad (\mathbf{14.1.f})$$

To be sure that the equation **(14.1.e)** is relativistic must be satisfied the energy-impulse relationship:

$$\left(\frac{E}{c}\right)^2 - p_1^2 - p_2^2 - p_3^2 = m^2 c^2$$

where the term $\left(\frac{E}{c}\right)^2$ is the squared of the first component $p_0^2$. To such a purpose let's multiply both members of equation **(14.1.d')** times the operator $(p_0 - \alpha_1 p_1 - \alpha_2 p_2 - \alpha_3 p_3 - \beta mc)$ and, since $\alpha_k$ and $\beta$ are matrices, we must have the shrewdness to respect the multiplication order. Moreover, since we supposed that $\alpha_k$ and $\beta$ represent a new degree of freedom, these matrices must be hermitian. Performing the product we get:

$$(p_0 + \alpha_1 p_1 + \alpha_2 p_2 + \alpha_3 p_3 + \beta mc) \cdot (p_0 - \alpha_1 p_1 - \alpha_2 p_2 - \alpha_3 p_3 - \beta mc)|\psi\rangle = 0$$

$$\{p_0^2 - [\alpha_1 p_1^2 + \alpha_2 p_2^2 + \alpha_3 p_3^2 + (\alpha_1\alpha_2 + \alpha_2\alpha_1)p_1 p_2 + (\alpha_1\alpha_3 + \alpha_3\alpha_1)p_1 p_3$$
$$+ (\alpha_2\alpha_3 + \alpha_3\alpha_2)p_2 p_3$$
$$+ (\alpha_1\beta + \beta\alpha_1)p_1 mc + (\alpha_2\beta + \beta\alpha_2)p_2 mc + (\alpha_3\beta + \beta\alpha_3)p_3 mc] - \beta^2 m^2 c^2\}|\psi\rangle$$
$$= 0 \qquad (\mathbf{14.1.g})$$

The (**14.1.g**) represents a system of differential equations of the second order that is reduced to the Klein-Gordon equation if the matrices $\alpha_k$ and $\beta$ satisfy the following relationship:

$$\begin{cases} (\alpha_k\alpha_l + \alpha_l\alpha_k) = 2\delta_{kl} \\ \alpha_k\beta + \beta\alpha_k = 0 \qquad (\mathbf{14.1.h}) \\ (\alpha_k)^2 = \beta^2 = \mathbb{1} \end{cases}$$

where the indices $k$ and $l$ run over all the permutations appearing in the (**14.1.g**). The relations (**14.1.h**) show that the matrices $\alpha_k$ and $\beta$ anti-commute and their squared is the unit matrix. Before to proceed in our discussion we need to clarify that the eigenfunctions $|\psi\rangle$ solutions of the equation (**14.1.d**) are also solutions of the (**14.1.g**), but the inverse is not always true. Dirac obtained the final explicit form of these equations using the Pauli matrices $\sigma_x, \sigma_y, \sigma_z$ for the spin operators:

$$\alpha_1 = \begin{pmatrix} 0 & 0 & 0 & 1 \\ 0 & 0 & 1 & 0 \\ 0 & 1 & 0 & 0 \\ 1 & 0 & 0 & 0 \end{pmatrix} , \ \alpha_2 = \begin{pmatrix} 0 & 0 & 0 & -i \\ 0 & 0 & i & 0 \\ 0 & -i & 0 & 0 \\ i & 0 & 0 & 0 \end{pmatrix} , \ \alpha_3 = \begin{pmatrix} 0 & 0 & 1 & 0 \\ 0 & 0 & 0 & -1 \\ 1 & 0 & 0 & 0 \\ 0 & -1 & 0 & 0 \end{pmatrix}$$

$$\beta = \begin{pmatrix} 1 & 0 & 0 & 0 \\ 0 & 1 & 0 & 0 \\ 0 & 0 & -1 & 0 \\ 0 & 0 & 0 & -1 \end{pmatrix}$$

These matrices are all hermitian and verify the properties (**14.1.h**). Therefore, the Dirac approach allows getting the electron spin without the necessity to introduce any new hypothesis. It follows that the spin is a physical property of the electron connected to its relativistic behavior. But the most part of atomic electrons have velocities lower than that of light, nevertheless they have always a spin. As a matter of fact the nature of spin is still unknown; the spin is an intrinsic property of the electron (and of the other elementary particles) and this means that it is independent by its velocity.

$\alpha_k$ and $\beta$ are of 4x4 matrices and this is due to the fact that the relativistic space has four dimensions. Therefore, the (**14.1.d**) is equivalent to a system of four differential equations, each of which allows to calculate one of the four components of the vector $|\psi\rangle$.

Equation (**14.1.d**) can be simplified multiplying both the members times the $\beta$ matrix:

$$\left( \frac{1}{c}\beta \frac{\partial}{\partial t} + \sum_{k=1}^{3} \alpha_k \beta \frac{\partial}{\partial x^k} + \frac{mc\beta^2}{i\hbar} \right) |\psi\rangle = 0$$

Recalling that $\beta^2 = \mathbb{1}$ and setting $\gamma^0 = \beta$ and $\alpha_k \beta = \gamma^k$ we get:

$$\left( \frac{1}{c}\gamma^0 \frac{\partial}{\partial t} + \sum_{k=1}^{3} \gamma^k \frac{\partial}{\partial x^k} + \frac{mc}{i\hbar}\mathbb{1} \right) |\psi\rangle = 0 \qquad (\mathbf{14.1.}\,\boldsymbol{i})$$

We note that while the matrix $\gamma^0$ is hermitian those $\gamma^k$ are anti-hermitian:

$$(\gamma^k)^\dagger = -\gamma^k$$

$$(\gamma^k)^2 = -\mathbb{1}$$

Moreover, these matrices satisfy the following commutation relations:

$$\gamma^\mu \gamma^\nu + \gamma^\nu \gamma^\mu = 2g^{\mu\nu}\mathbb{1}$$

Equation (**14.1.i**) can be then rewritten as:

$$\left( \sum_{\mu=0}^{3} i\hbar\gamma^\mu \frac{\partial}{\partial x^\mu} - mc\,\mathbb{1} \right) |\psi\rangle = 0$$

Denoting the operator $\frac{\partial}{\partial x^\mu}$ by the symbol $\partial_\mu$ we arrive to the final and compact form of Dirac equation:

$$\left( \gamma^\mu p_\mu - mc\mathbb{1} \right)|\psi\rangle = \left( \gamma_\mu p^\mu - mc\mathbb{1} \right)|\psi\rangle = 0 \qquad (\mathbf{14.1.}\,\boldsymbol{l})$$

where both tensor product (contravariant components of $\gamma$ times covariant components of $p$ and vice versa) have been shown. We proved that the eigenfunction $|\psi\rangle$ is a four-vector whose components could be complex. We can so define the probability density function $\rho$ as:

$$\rho = \langle\psi|\psi\rangle = \psi^\dagger \psi$$

In such a way the value of the function $\rho$ is always defined and positive. Moreover, the probability flow $\boldsymbol{J}$ will have four components given by:

$$J^k = c\psi^\dagger \alpha^k \psi \qquad (\mathbf{14.1.}\,\boldsymbol{m})$$

The components of this vector may be written using the matrices $\gamma^k$ instead of $\alpha^k$; to do this we recall that:

$$\gamma^k = \beta\alpha^k$$

that if multiplied times $\beta$ in both side give us:

$$\beta\gamma^k = \beta^2 \alpha^k = \alpha^k$$

Replacing this result in (**14.1.m**) we get:

$$J^k = c\psi^\dagger \beta\gamma^k \psi = c\psi^\dagger \gamma^0 \gamma^k \psi = \bar{\psi}\gamma^k \psi \qquad (\mathbf{14.1.}\,\boldsymbol{n})$$

where in the second member $k = 1, 2, 3$ , $\beta = \gamma^0$ and $\bar{\psi}$ is simply the complex conjugated of the vector $\psi$.

Equation (**14.1.n**) is symmetric respect to the space-time coordinates and this is the condition required to be relativistic. But we have to prove that it is invariant respect the Lorentz transformations. That means proving that Dirac matrices remain unchanged under Lorentz transformations. The transformed vector $|\psi\rangle$, which is denoted by ψ', must be the linear transformed of ψ:

$$\psi' = S\psi$$

and the transformation matrix $S$ must satisfy the condition:

$$S^{-1}\gamma^k S = \Lambda_\mu^k \gamma^\mu$$

where $\Lambda_\mu^k$ is the tensor representing the Lorentz transformations:

$$\Lambda_\mu^k = \begin{pmatrix} \Gamma & \Gamma B & 0 & 0 \\ \Gamma B & \Gamma & 0 & 0 \\ 0 & 0 & 1 & 0 \\ 0 & 0 & 0 & 1 \end{pmatrix}$$

Here $\Gamma = (1 - B^2)^{-1/2}$ and $B = v/c$ where $v$ is the relative velocity of the framework $K'$ respect the framework $K$ one. This matrix transforms the space-time coordinates passing from a framework to another one:

$$x^k = \Lambda_\mu^k x^\mu$$

The complex conjugated of the vector $|\psi\rangle$ will be transformed in a symmetrical mode:

$$\overline{\psi'} = \bar{\psi} S^{-1}$$

We can now evaluate the behavior of the probability flow under the Lorentz transformations:

$$(J^k)' = \overline{\psi'}\gamma^k \psi' = \bar{\psi} S^{-1}\gamma^k S\psi = \bar{\psi}\Lambda_\mu^k \gamma^\mu \psi = \Lambda_\mu^k \bar{\psi}\gamma^\mu \psi = \Lambda_\mu^k J^\mu$$

which proves its invariance under the Lorentz transformations. We proved that Dirac equation for the free electron is relativistically invariant.

We go back to the previous discussion about the dimension of vector $|\psi\rangle$; it has four components arising from the fact that the Dirac matrices are 4x4. However, we know that the electron spin has only two quantized components. Why then $|\psi\rangle$ must have four components instead of two? Are there any components of the vector $|\psi\rangle$ physically not acceptable? The answers will be given in the next section, where will be shown how the Dirac equation was able to predict the existence of a new particle (positron) whose discovery was done some years later the publication of the Dirac theory.

The four-vector $|\psi\rangle$ is called spinor since is a vector containing the information about the half-integer spin of the particle; however it must not be confused as a vector of the Minkowski space because it does not transform according to the Lorentz transformation. For that reason the Dirac equation usually is called as spinor equation.

## 14.2. The Free Relativistic Electron

Equation **(14.1.d)** describes the relativistic behavior of a free electron. The solutions are plane waves whose components are given by:

$$\psi_j = u_j(\boldsymbol{p}) exp\left\{-i\frac{p \cdot x}{\hbar}\right\} \quad 0 \le j \le 3$$

Replacing the column vector $|\psi\rangle$ in equation **(14.1.d)** and using the explicit form of the matrices $\alpha_k$ and $\beta$, we get a system of four linear differential equations where the unknown quantities are the functions $u_j(\boldsymbol{p})$:

$$\left[-i\frac{\hbar}{c}\frac{\partial}{\partial t}\begin{pmatrix}1&0&0&0\\0&1&0&0\\0&0&1&0\\0&0&0&1\end{pmatrix} - i\hbar\frac{\partial}{\partial x_1}\begin{pmatrix}0&0&0&1\\0&0&1&0\\0&1&0&0\\1&0&0&0\end{pmatrix} - i\hbar\frac{\partial}{\partial x_2}\begin{pmatrix}0&0&0&-i\\0&0&i&0\\0&-i&0&0\\i&0&0&0\end{pmatrix} - \right.$$

$$\left. -i\hbar\frac{\partial}{\partial x_3}\begin{pmatrix}0&0&1&0\\0&0&0&-1\\1&0&0&0\\0&-1&0&0\end{pmatrix} + mc\begin{pmatrix}1&0&0&0\\0&1&0&0\\0&0&-1&0\\0&0&0&-1\end{pmatrix}\right]\begin{pmatrix}u_0(\boldsymbol{p})e^{-i\frac{p\cdot x}{\hbar}}\\u_1(\boldsymbol{p})e^{-i\frac{p\cdot x}{\hbar}}\\u_2(\boldsymbol{p})e^{-i\frac{p\cdot x}{\hbar}}\\u_3(\boldsymbol{p})e^{-i\frac{p\cdot x}{\hbar}}\end{pmatrix} = \begin{pmatrix}0\\0\\0\\0\end{pmatrix}$$

$$\begin{bmatrix}-i\hbar\partial_{x_0} & 0 & -i\hbar\partial_{x_3} & (-i\hbar\partial_{x_1}-i\hbar\partial_{x_2})\\0 & (-i\hbar\partial_{x_0}+mc) & (-i\hbar\partial_{x_1}+i\hbar\partial_{x_2}) & i\hbar\partial_{x_3}\\-i\hbar\partial_{x_3} & (-i\hbar\partial_0-i\hbar\partial_{x_2}) & (-i\hbar\partial_{x_0}-mc) & 0\\(-i\hbar\partial_{x_1}+i\hbar\partial_{x_2}) & i\hbar\partial_{x_3} & 0 & (-i\hbar\partial_{x_0}-mc)\end{bmatrix}\begin{bmatrix}u_0(\boldsymbol{p})\\u_1(\boldsymbol{p})\\u_2(\boldsymbol{p})\\u_3(\boldsymbol{p})\end{bmatrix} = \begin{bmatrix}0\\0\\0\\0\end{bmatrix}$$

$$\begin{cases}-i\hbar\partial_{x_0}u_0(\boldsymbol{p})--i\hbar\partial_{x_3}u_{2(\boldsymbol{p})}+(-i\hbar\partial_{x_1}-i\hbar\partial_{x_2})u_3(\boldsymbol{p})=0\\(-i\hbar\partial_{x_0}+mc)u_1(\boldsymbol{p})+(-i\hbar\partial_{x_1}+i\hbar\partial_{x_2})u_2(\boldsymbol{p})+i\hbar\partial_{x_3}u_3(\boldsymbol{p})=0\\-i\hbar\partial_{x_3}u_0(\boldsymbol{p})+(-i\hbar\partial_0-i\hbar\partial_{x_2})u_1(\boldsymbol{p})+(-i\hbar\partial_{x_0}-mc)u_2(\boldsymbol{p})=0\\(-i\hbar\partial_{x_1}+i\hbar\partial_{x_2})u_0(\boldsymbol{p})+i\hbar\partial_{x_3}u_1(\boldsymbol{p})+(-i\hbar\partial_{x_0}-mc)u_3(\boldsymbol{p})=0\end{cases} \quad (\mathbf{14.2.\boldsymbol{a}})$$

where the partial derivatives performed respect the space-time coordinates have been denoted by $\partial_{x_i}$. The linear system **(14.2.a)** has non-trivial solutions if the determinant of the coefficients matrix is zero. The value of this determinant is:

$$(E^2 - p^2c^2 - m^2c^4) = 0 \qquad (\mathbf{14.2.\boldsymbol{b}})$$

that is the squared of the relativistic energy. The roots of **(14.2.b)** are:

$$E = \pm\sqrt{p^2c^2 + m^2c^4}$$

As expected, the energy of the free electron is not quantized since the impulse $p$ may assume any value. However, the relativistic approach proves that the energy can assume also negative values;

this result is the novelty introduced by the Dirac equation. Let's suppose $u_+(\boldsymbol{p})$ is a solution of the system (**14.2.a**) with positive energy. The relativistic Hamiltonian operator is given by (**14.1.f**) that in compact form can be written as:

$$H = c\alpha \cdot \frac{\hbar}{i}\nabla + \beta mc^2 = c\alpha \cdot \boldsymbol{p} + \beta mc^2$$

Since $|\psi\rangle$ is a solution of equation (**14.1.d**) then the function $u_+(\boldsymbol{p})$ is an eigenvector of the Hamiltonian operator:

$$(c\alpha \cdot \boldsymbol{p} + \beta mc^2)u_+(\boldsymbol{p}) = E(\boldsymbol{p})u_+(\boldsymbol{p}) \qquad (\boldsymbol{14.2.c})$$

This eigenvector can be written as:

$$u_+(\boldsymbol{p}) = \begin{pmatrix} u_1 \\ u_2 \end{pmatrix}$$

The (**14.2.c**) becomes then a system of two differential equations:

$$\begin{cases} c\sigma \cdot \boldsymbol{p}u_2 + mc^2 u_1 = E(\boldsymbol{p})u_1 \\ c\sigma \cdot \boldsymbol{p}u_1 - mc^2 u_2 = E(\boldsymbol{p})u_2 \end{cases} \qquad (\boldsymbol{14.2.d})$$

where $\sigma$ are the Pauli matrices. From the second equation of the (**14.2.d**) we get:

$$u_2 = c\frac{\sigma \cdot \boldsymbol{p}}{E(\boldsymbol{p}) + mc^2}u_1 \qquad (\boldsymbol{14.2.e})$$

Therefore, the arbitrary setting of $u_1$ allows to obtain the correspondent function $u_2$. The system (**14.2.d**) has linear independent solutions for each value of the impulse $\boldsymbol{p}$. Setting $u_1$ as:

$$u_1 = \begin{pmatrix} 1 \\ 0 \end{pmatrix} \quad or \quad u_1 = \begin{pmatrix} 0 \\ 1 \end{pmatrix}$$

we get the explicit form of the eigenvectors:

$$u_+^{(1)}(\boldsymbol{p}) = \begin{pmatrix} 1 \\ 0 \\ c\dfrac{\sigma \cdot \boldsymbol{p}}{E(\boldsymbol{p}) + mc^2} \\ 0 \end{pmatrix} \quad , \quad u_+^{(2)}(\boldsymbol{p}) = \begin{pmatrix} 1 \\ 0 \\ 0 \\ c\dfrac{\sigma \cdot \boldsymbol{p}}{E(\boldsymbol{p}) + mc^2} \end{pmatrix}$$

The two vectors must be normalized using the condition $u^*u = 1$. In the non-relativistic limit $E(\boldsymbol{p})$ is small compare to $mc^2$ so that the vector component $c\frac{\sigma \cdot \boldsymbol{p}}{E(\boldsymbol{p}) + mc^2}$ can be approximated to $\frac{\sigma \cdot \boldsymbol{p}}{mc}$ and the (**14.2.e**) becomes:

$$u_2 = \frac{\sigma \cdot \boldsymbol{p}}{mc}u_1$$

Replacing the component $u_2$ in the first of equations (**14.2.d**) we get:

$$c(\sigma \cdot \boldsymbol{p})\frac{\sigma \cdot \boldsymbol{p}}{2mc}u_1 + mc^2 u_1 = E(\boldsymbol{p})u_1$$

$$\left[\frac{(\sigma \cdot \boldsymbol{p})^2}{2m} + mc^2 - E(\boldsymbol{p})\right]u_1 = 0$$

This is the Schrödinger equation for the free electron where the relativistic term $mc^2$ has been added:

$$\left[ -\frac{\hbar^2}{2m}\nabla^2 + mc^2 \right] u_1 = E(\boldsymbol{p}) u_1$$

In fact, the term $\sigma \cdot \boldsymbol{p}$ corresponds to the following matrix sum:

$$\sum_{k=1}^{3} \sigma_k p^k = -i\hbar \sum_{k=1}^{3} \sigma_k \frac{\partial}{\partial x^k} = -i\hbar \sum_{k=1}^{3} \sigma_k \partial_{x^k}$$

Replacing the symbols with the explicit forms of the Pauli matrices we obtain:

$$-i\hbar \begin{pmatrix} 0 & 1 \\ 1 & 0 \end{pmatrix} \partial_x - i\hbar \begin{pmatrix} 0 & -i \\ i & 0 \end{pmatrix} \partial_y - i\hbar \begin{pmatrix} 1 & 0 \\ 0 & -1 \end{pmatrix} \partial_z = \begin{pmatrix} -i\hbar \partial_z & i\hbar \partial_y - i\hbar \partial_x \\ -i\hbar \partial_x - i\hbar \partial_y & i\hbar \partial_z \end{pmatrix}$$

The term $(\sigma \cdot \boldsymbol{p})^2$ becomes then a diagonal matrix whose non-vanishing elements are $\nabla^2$; this proves the form of the kinetic energy operator above written in the Schrödinger equation. The two eigenfunctions $|\psi\rangle$ solutions of Dirac equation for the free electron with positive energy are:

$$\begin{cases} |\psi\rangle^{(1)} = u_+^{(1)}(\boldsymbol{p}) exp\left\{ -i\hbar \dfrac{\boldsymbol{p} \cdot \boldsymbol{x}}{\hbar} \right\} \\ |\psi\rangle^{(2)} = u_+^{(2)}(\boldsymbol{p}) exp\left\{ -i\hbar \dfrac{\boldsymbol{p} \cdot \boldsymbol{x}}{\hbar} \right\} \end{cases} \qquad (\boldsymbol{14.2.f})$$

which differ for the spin state.

Let's consider now the solutions with negative energy; it easy to verify that the Dirac equation has two linear independent solutions similar to the **(14.2.f)** with different spin. To simply the interpretation of these solutions we consider the non-relativistic limit where the energy is close to $-mc^2$. By this approximation we get:

$$u_2 = -\frac{\sigma \cdot \boldsymbol{p}}{mc} u_1$$

where $u_1$ and $u_2$ are the components of the vector $u_-(\boldsymbol{p})$. The Schrödinger equation then becomes:

$$\left[ \frac{\hbar^2}{2m}\nabla^2 + mc^2 \right] u_1 = -|E(\boldsymbol{p})| u_1$$

where the sign of the mass is negative. Then, we can state that for the relativistic electron two states with negative kinetic energy, negative mass and opposite spin are allowed; they obey to the Pauli exclusion principle. These *unusual* states have physical meaning because we can prove that the probability transition from a state to negative energy to another with positive energy is not zero. To explain the states with negative energy Dirac proposed the *hole theory* model, according which all the states with negative energy are occupied according to the exclusion principle. In an unperturbed situation is therefore impossible that a transition from a negative to a positive state occurs. That means the states having negative energy remain *invisible*! Perturbing by an external field these latent states is possible take out from one of them a particle making so *visible* the *new* physical *object* having the same mass of the electron but positive charge. This particle postulated by Dirac in 1928 is what that today we call positron and it was experimentally discovered in 1932 in the cosmic ray by the physician Carl Anderson [30].

The most general solution of Dirac equation for the free electron is the sum of all its solutions, both with positive and negative energies (relativistic wave packet). Therefore, if we perform a measure of position on a relativistic electron (using high energy photons), a couple electron-positron could be created: this phenomenon is completely fortuitous and cannot be controlled during the measurement. So, the question is: of which particle are we measuring the position? It's quite easy to understand that it has not more sense speaking of quantum particle position. A relativistic quantum system has infinite degrees of freedom, like the photons in the electromagnetic field, and their numbers may change continuously. We understand now that also the law of conservation of mass loses its meaning and only the law of energy conservation keeps up its validity. In the modern relativistic quantum theory the mathematical formalism is that of the creator-annihilator operators that we already encountered in the matrix mechanics theory.

Finally, we note that since the Dirac equation is relativistic the Hamiltonian operator depends by the time. We are so in the Heisenberg picture and the whole relativistic quantum mechanics is developed according this formalism.

## 14.3. The Electronic Spin

The electron spin is one of the most important discoveries in quantum mechanics and allows explaining many properties of atoms and molecules. The physicist that most contributed to the study of the spin has been Pauli that, during the first half of the last century, formulated what today we know as exclusion principle [31].

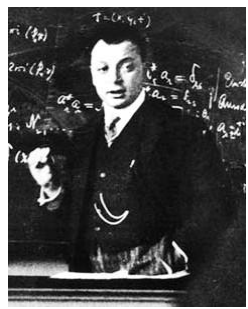

**Wolfgang Pauli**

How previously proved, the spin is a degree of freedom due to the relativistic behavior of the electron; in principle it may be associated to a sort of motion of the particle described by an unusual angular momentum called spin. This quantity is a constant of the motion and for that it must verify the following commutation relation:

$$[H, \hat{S}] = 0$$

Let's consider the angular momentum operator associated to the electron:

$$\hat{L} = \boldsymbol{r} \times \hat{p} = -i\hbar \boldsymbol{r} \times \nabla$$

If we assume that the Hamiltonian of the relativistic free electron is:

$$H = -i\hbar c \alpha \nabla + \beta mc^2$$

it's easy to verify that the commutator $[H, \hat{S}]$ is different from zero and that the angular momentum is not a constant of its motion. But if we take the sum between the operator $\hat{L}$ and $\hat{S} = \frac{1}{2}\hbar\sigma$ we get the spin-orbit coupling operator:

$$\hat{J} = \hat{L} + \frac{1}{2}\hbar\sigma$$

This operator commutes with the relativistic Hamiltonian proving that it is a constant of motion. Therefore, Dirac equation **(14.1.d)** describes the motion of free particles with half-integer spin $\frac{1}{2}$.

We know that the spin components along the z axis may assume two discrete values given by $\frac{\hbar}{2}$ and $-\frac{\hbar}{2}$. The spin states for the electron and positron are:

$$\left|\frac{1}{2}, \frac{1}{2}\right\rangle \quad , \quad \left|\frac{1}{2}, -\frac{1}{2}\right\rangle$$

and the general state of the particle is:

$$|\chi\rangle = c_1 \left|\frac{1}{2}, \frac{1}{2}\right\rangle + c_2 \left|\frac{1}{2}, -\frac{1}{2}\right\rangle$$

where $c_1$ and $c_2$ are complex coefficients whose squared modulus gives the probability that the particles have a z-component of $\frac{\hbar}{2}$ or $-\frac{\hbar}{2}$.

The application of a constant magnetic field produces the splitting of two states of spin (Zeeman effect). In this case the Hamiltonian operator of the particle is:

$$H = -\frac{\hbar^2}{2m}\left(\nabla - i\hbar\frac{e}{c}\boldsymbol{A}\right)^2 - eV - 2\mu_B\hat{S} \cdot \frac{\boldsymbol{B}}{\hbar} \qquad (\boldsymbol{14.3.a})$$

where $V$ is the potential energy of the particle, $\boldsymbol{B}$ is the magnetic field and $\boldsymbol{A}$ is the potential vector satisfying the condition:

$$\boldsymbol{B} = \nabla \times (\boldsymbol{A} + \nabla \Phi)$$

with $\boldsymbol{\Phi}$ scalar field. Making explicit the squared of the **(14.3.a)** and neglecting the squared potential vector we obtain:

$$H = -\frac{\hbar^2}{2m}\nabla^2 - eV - \mu_B \hat{L} \cdot \frac{\boldsymbol{B}}{\hbar} - 2\mu_B \hat{S} \cdot \frac{\boldsymbol{B}}{\hbar} \qquad (\boldsymbol{14.3.b})$$

where the number 2 in the last operatorial term is the symmetry factor $g_s$ deduced by experimental measurements. Recalling the time dependent Schrödinger equation and considering that the eigenfunction is a two components vector (we are considering a non- relativistic case) we arrive to the Pauli equation:

$$i\hbar \frac{\partial}{\partial t}\begin{pmatrix}\psi_+\\\psi_-\end{pmatrix} = \left(-\frac{\hbar^2}{2m}\nabla^2 - eV - \mu_B \hat{L} \cdot \frac{\boldsymbol{B}}{\hbar} - 2\mu_B \hat{S} \cdot \frac{\boldsymbol{B}}{\hbar}\right)\begin{pmatrix}\psi_+\\\psi_-\end{pmatrix} \qquad (\boldsymbol{14.3.c})$$

This equation is based on the experimental results and does not include the operatorial term of the the spin-orbit coupling.

We want to prove now that the same result may be obtained starting from the relativistic Hamiltonian without the necessity to use corrections *ad hoc* to fit the theory to the experimental results. To such a purpose we replace in the relativistic Hamiltonian $H = c\alpha \cdot \boldsymbol{p} + \beta mc^2$ the impulse $\boldsymbol{p}$ by a four-vector with components $p^k - \frac{e}{c}A^k$ (this replacing is called minimal substitution):

$$H = c\sum_{k=1}^{3}\alpha_k\left(p^k - \frac{e}{c}A^k\right) - eV + \beta mc^2 = c\sum_{k=1}^{3}\alpha_k p^k + \beta mc^2 + c\sum_{k=1}^{3}\alpha_k\left(-\frac{e}{c}A^k\right) - eV =$$

$$= H_0 - e\sum_{k=1}^{3}\alpha_k A^k - eV = H_0 - e\sigma \cdot \boldsymbol{A} - eV$$

where $H_0$ is the relativistic unperturbed Hamiltonian operator and $V$ is the electric scalar potential (we recall that the particle is interacting with an external electromagnetic field). Then, Dirac equation can be written as:

$$i\hbar \frac{\partial}{\partial t}|\psi\rangle = \left[c\alpha \cdot \left(\mathbf{p} - \frac{e}{c}\boldsymbol{A}\right) + \beta mc^2 - eV\right]|\psi\rangle$$

Assuming that the vector $|\psi\rangle$ is formed by two components $\varphi'$ and $\chi'$, setting $\left(\mathbf{p} - \frac{e}{c}\boldsymbol{A}\right) = \boldsymbol{\pi}$, using the explicit form of the matrices $\alpha_k$ and $\beta$ we arrive to the equation:

$$i\hbar \frac{\partial}{\partial t}\begin{pmatrix}\varphi'\\\chi'\end{pmatrix} = c\sigma \cdot \boldsymbol{\pi}\begin{pmatrix}\varphi'\\\chi'\end{pmatrix} + mc^2\begin{pmatrix}\varphi'\\\chi'\end{pmatrix} - eV\begin{pmatrix}\varphi'\\\chi'\end{pmatrix} \qquad (\boldsymbol{14.3.d})$$

In the Schrödinger picture the time evolution of the eigenfunction is given by $exp\left\{-i\frac{Et}{\hbar}\right\}$ and in case of the relativistic free electron becomes $exp\left\{-i\frac{(mc^2+T)t}{\hbar}\right\}$ with $T$ kinetic energy. In the non-relativistic limit the energy term due to the mass prevails over the others and then is possible write:

$$\begin{pmatrix} \varphi' \\ \chi' \end{pmatrix} = exp\left\{-i\frac{(mc^2 + T)t}{\hbar}\right\}\begin{pmatrix} \varphi \\ \chi \end{pmatrix} \cong exp\left\{-i\frac{mc^2 t}{\hbar}\right\}\begin{pmatrix} \varphi \\ \chi \end{pmatrix} \qquad (\mathbf{14.3.e})$$

where $\varphi$ and $\chi$ are vectorial functions which depend only on the space and spin coordinates. Replacing the approximation (**14.3.e**) in the (**14.3.d**) we get:

$$i\hbar\frac{\partial}{\partial t}exp\left\{-i\frac{mc^2 t}{\hbar}\right\}\begin{pmatrix} \varphi \\ \chi \end{pmatrix} = \left\{c\sigma\cdot\boldsymbol{\pi}\begin{pmatrix} \varphi \\ \chi \end{pmatrix} + mc^2\begin{pmatrix} \varphi \\ \chi \end{pmatrix} - eV\begin{pmatrix} \varphi \\ \chi \end{pmatrix}\right\}exp\left\{-i\frac{mc^2 t}{\hbar}\right\}$$

$$i\hbar\frac{\partial}{\partial t}\begin{pmatrix} \varphi \\ \chi \end{pmatrix} + mc^2\begin{pmatrix} \varphi \\ \chi \end{pmatrix} = c\sigma\cdot\boldsymbol{\pi}\begin{pmatrix} \chi \\ \varphi \end{pmatrix} + mc^2\begin{pmatrix} \varphi \\ -\chi \end{pmatrix} - eV\begin{pmatrix} \varphi \\ \chi \end{pmatrix}$$

$$i\hbar\frac{\partial}{\partial t}\begin{pmatrix} \varphi \\ \chi \end{pmatrix} = c\sigma\cdot\boldsymbol{\pi}\begin{pmatrix} \chi \\ \varphi \end{pmatrix} + mc^2\left[\begin{pmatrix} \varphi \\ -\chi \end{pmatrix} - \begin{pmatrix} \varphi \\ \chi \end{pmatrix}\right] - eV\begin{pmatrix} \varphi \\ \chi \end{pmatrix}$$

$$i\hbar\frac{\partial}{\partial t}\begin{pmatrix} \varphi \\ \chi \end{pmatrix} = c\sigma\cdot\boldsymbol{\pi}\begin{pmatrix} \chi \\ \varphi \end{pmatrix} - 2mc^2\left[\begin{pmatrix} 0 \\ \chi \end{pmatrix}\right] - eV\begin{pmatrix} \varphi \\ \chi \end{pmatrix} \qquad (\mathbf{14.3.f})$$

The (**14.3.f**) is a system of two coupled differential equations. We focus our attention on the second one:

$$i\hbar\frac{\partial}{\partial t}\chi = c\sigma\cdot\boldsymbol{\pi}\varphi - 2mc^2\chi - eV\chi \qquad (\mathbf{14.3.g})$$

If we suppose that the function $\chi$ changes slowly over the time, which implies $\frac{\partial\chi}{\partial t} = 0$, and that the interaction of the particle with the electric potential $V$ is negligible (the electric field of an electromagnetic wave usually is quite small compared to the field generated by the same electrical particle), then the (**14.3.g**) becomes:

$$c\sigma\cdot\boldsymbol{\pi}\varphi - 2mc^2\chi = 0$$

by which we get:

$$\chi = \frac{\sigma\cdot\boldsymbol{\pi}}{2mc}\varphi$$

Replacing this result in the first equation of (**14.3.f**):

$$i\hbar\frac{\partial}{\partial t}\varphi = c\sigma\cdot\boldsymbol{\pi}\frac{\sigma\cdot\boldsymbol{\pi}}{2mc}\varphi - eV\varphi$$

$$i\hbar\frac{\partial}{\partial t}\varphi = \frac{1}{2m}(\sigma\cdot\boldsymbol{\pi})^2\varphi - eV\varphi$$

$$i\hbar\frac{\partial}{\partial t}\varphi = \frac{1}{2m}\left[\sigma\cdot\left(\mathbf{p} - \frac{e}{c}\boldsymbol{A}\right)\right]^2\varphi - eV\varphi$$

We calculate now the product $(\sigma\cdot\boldsymbol{\pi})^2$:

$$\sigma\cdot\boldsymbol{\pi}\sigma\cdot\boldsymbol{\pi} = \boldsymbol{\pi}\cdot\boldsymbol{\pi} + i\sigma(\boldsymbol{\pi}\times\boldsymbol{\pi})$$

Replacing $\mathbf{p}$ with the operator $-i\hbar\nabla$ and using the last product we get:

$$i\hbar\frac{\partial}{\partial t}\varphi = \left[\frac{1}{2m}\left(\mathbf{p} - \frac{e}{c}\boldsymbol{A}\right)^2 - \frac{e\hbar}{2mc}\sigma\cdot\boldsymbol{B} - eV\right]\varphi \qquad (\mathbf{14.3.h})$$

where have been neglected the squared terms. Recalling that the Bohr magneton is $\mu_B = \frac{e\hbar}{2mc}$ the (14.3.h) can be written as:

$$i\hbar \frac{\partial}{\partial t} \varphi = \left[ \frac{1}{2m} \left( \mathbf{p} - \frac{e}{c} \boldsymbol{A} \right)^2 - \mu_B \sigma \cdot \boldsymbol{B} - eV \right] \varphi$$

$$i\hbar \frac{\partial}{\partial t} \varphi = \left[ \frac{1}{2m} \left( \mathbf{p} - \frac{e}{c} \boldsymbol{A} \right)^2 - 2\mu_B \hat{S} \cdot \frac{\boldsymbol{B}}{\hbar} - eV \right] \varphi \qquad (\mathbf{14.3.i})$$

The (14.3.i) is similar to the Pauli equation (14.3.c); $\varphi$ is a column vector with two components which differ for the spin coordinate. In equation (14.3.i) appears the symmetry factor $g_s = 2$ without the necessity to have inserted it *ad hoc*. We conclude that the Dirac equation includes that of Pauli which describes the particles with half-integer spin. In the case the magnetic field is weak and uniform the (14.3.i) becomes:

$$i\hbar \frac{\partial}{\partial t} \varphi = \left[ -\frac{\hbar^2}{2m} \nabla^2 - \frac{e}{2mc} \left( \hat{L} + 2\hat{S} \right) \cdot \boldsymbol{B} \right] \varphi$$

where explicitly appears the spin-orbit coupling operator.

## 14.4 The Dirac Theory for the Hydrogen Atom

By the use of the explicit form of the electron potential energy in hydrogen atom, $V(r) = -e^2/4\pi\varepsilon_0 r$; the (14.1.d) can be written as:

$$\begin{cases} (E - V(r) - mc^2)\varphi = -i\hbar c\sigma \cdot \nabla\chi \\ (E - V(r) + mc^2)\chi = -i\hbar c\sigma \cdot \nabla\varphi \end{cases} \qquad (\mathbf{14.4.a})$$

where the functions $\varphi$ and $\chi$ are column vectors with two components. Taking from the second equation the function $\chi$ and developing it as power series respect the variable $(E - V(r) - mc^2)^{-1}$ we obtain:

$$\chi = -\frac{i\hbar}{2mc} \left( 1 - \frac{E - V(r) - mc^2}{2mc^2} + \cdots \right) \sigma \cdot \nabla\varphi$$

Since the term $(E - V(r) - mc^2)$ is small respect $2mc^2$, we can cut the development to the second term of the summation. Replacing the approximated vectorial function $\chi$ in the first equation of (14.4.a) we get:

$$\left[ E - V(r) + \frac{\hbar^2}{2m} \nabla^2 + \frac{\hbar^2}{(2mc)^2} \left( \sigma \cdot \nabla V(r) \right) \sigma \cdot \nabla + \frac{\hbar^2}{(2mc)^2} (E - V(r)) \nabla^2 \right] \varphi = 0$$

The term in the square bracket could be seen as:

$$\frac{\hbar^2}{(2mc)^2} (E - V) \nabla^2 = \frac{\hbar^2}{(2mc)^2} \left( \nabla^2 V + 2\nabla V \cdot \nabla + \nabla^2 (E - V) \right)$$

and since $(E - V)$ is the kinetic energy of the particle, associated to the operator $-\frac{\hbar^2}{2m}\nabla^2$, the last operator becomes:

$$\frac{\hbar^2}{(2mc)^2}(E - V)\nabla^2 = \frac{\hbar^2}{(2mc)^2}\left(\nabla^2 V + 2\nabla V \cdot \nabla - \frac{\hbar^2}{2m}\nabla^4\right)$$

Replacing this operator in the previous equation we get:

$$\left[E - V + \frac{\hbar^2}{2m}\nabla^2 + (\sigma \cdot \nabla V)\sigma \cdot \nabla + \frac{\hbar^2}{(2mc)^2}\nabla^2 V + \frac{\hbar^2}{(2mc)^2}2\nabla V \cdot \nabla - \frac{\hbar^4}{8m^3c^2}\nabla^4\right]\varphi = 0$$

Considering that $\nabla^2 V = 2\pi e^2 \delta(r)/4\pi\varepsilon_0$, that the spin operator is $\hat{S}_k = \frac{1}{2}\hbar\sigma_k$, that $\sigma_k^2 = \mathbb{1}$, that $\sigma_h\sigma_k = i\sigma_l = \sigma_k\sigma_h$, that $\hat{L}_x, \hat{L}_y, \hat{L}_z$ are given by the **(14.1.b)**, we arrive to the final equation:

$$\left[-\frac{\hbar^2}{2m}\nabla^2 + V - \frac{\hbar^4}{8m^3c^2}\nabla^4 + \frac{1}{2m^2c^2}\frac{dV}{rdr}\hat{L}\cdot\hat{S} + \frac{\pi\hbar^2}{2m^2c^2}\frac{e^2}{4\pi\varepsilon_0}\delta(r)\right]\varphi = E\varphi \qquad (\mathbf{14.4.b})$$

The **(14.4.b)** is the Dirac equation for the hydrogen atom. The two first terms into square bracket represent the non-relativistic Hamiltonian $H_0$ whose solutions are $|n, l, m\rangle$. The term $-\frac{\hbar^4}{8m^3c^2}\nabla^4$ represents the relativistic correction of the kinetic energy of the electron. The term $\frac{1}{2m^2c^2}\frac{dV}{rdr}\hat{L}\cdot\hat{S}$ represents the spin-orbit interaction energy operator; its explicit form is:

$$\frac{1}{2m^2c^2}\frac{dV}{rdr}\hat{L}\cdot\hat{S} = \mu_B\frac{1}{mc^2e\hbar}\frac{dV}{rdr}\hat{L}\cdot\hat{S}$$

Once again is surprising how the prediction of this interaction is derived by a mathematical calculation and not by the need to fit the theory to the experimental results. Finally, the last operatorial term is the Darwin term and gives an energetic contribution only for the eigenfunctions whose probability density is non-zero on the core (that means eigenfunctions having quantum number $l = 0$).

The resolution of equation **(14.4.b)**, although approximated, involves considerable difficulties. Therefore, is convenient to take the non-relativistic solutions and calculate the mean expectation values of the three relativistic terms; that's the typical perturbation theory approach justified by the fact that the relativistic energy corrections are smaller than the Schrödinger energy. We start with the term $\frac{\hbar^4}{8m^3c^2}\nabla^4$ due to Thomas:

$$E_T = \left\langle n, l, m\left|\frac{\hbar^4}{8m^3c^2}\nabla^4\right|n, l, m\right\rangle = -\frac{1}{2mc^2}\left\langle n, l, m\left|\left(-\frac{\hbar^2}{2m}\nabla^2\right)^2\right|n, l, m\right\rangle$$

The term $-\frac{\hbar^2}{2m}\nabla^2$ is the kinetic energy operator and can be obtained by the non-relativistic Hamiltonian $H_0$:

$$H_0 = -\frac{\hbar^2}{2m}\nabla^2 - \frac{1}{4\pi\varepsilon_0}\frac{e^2}{r} \quad \rightarrow \quad -\frac{\hbar^2}{2m}\nabla^2 = H_0 + \frac{1}{4\pi\varepsilon_0}\frac{e^2}{r}$$

that replaced in the expression of $E_T$ gives:

$$E_T = -\frac{1}{2mc^2}\left\langle n,l,m\left|\left(H_0 + \frac{1}{4\pi\varepsilon_0}\frac{e^2}{r}\right)^2\right|n,l,m\right\rangle =$$

$$= -\frac{1}{2mc^2}\left\langle n,l,m\left|\left(H_0 + \frac{1}{4\pi\varepsilon_0}\frac{e^2}{r}\right)\left(H_0 + \frac{1}{4\pi\varepsilon_0}\frac{e^2}{r}\right)\right|n,l,m\right\rangle =$$

$$= -\frac{1}{2mc^2}\left\langle n,l,m\left|H_0{}^2 + H_0\frac{1}{4\pi\varepsilon_0}\frac{e^2}{r} + \frac{1}{4\pi\varepsilon_0}\frac{e^2}{r}H_0 + \left(\frac{e^2}{4\pi\varepsilon_0}\right)^2\frac{1}{r^2}\right|n,l,m\right\rangle =$$

$$= -\frac{1}{2mc^2}\left[E_n{}^2 + 2E_n\frac{e^2}{4\pi\varepsilon_0}\left\langle n,l,m\left|\frac{1}{r}\right|n,l,m\right\rangle + \left(\frac{e^2}{4\pi\varepsilon_0}\right)^2\left\langle n,l,m\left|\frac{1}{r^2}\right|n,l,m\right\rangle\right]$$

The results of the two integrals in the square bracket are:

$$\left\langle n,l,m\left|\frac{1}{r}\right|n,l,m\right\rangle = \frac{Z}{a_B n^2}$$

$$\left\langle n,l,m\left|\frac{1}{r^2}\right|n,l,m\right\rangle = \frac{Z^2}{a_B{}^2 n^3 l\left(l+\frac{1}{2}\right)}$$

Z is the atomic number that for the hydrogen atom is 1. Replacing these integrals in the last expression we arrive to the final form of the Thomas energy:

$$E_T = -\frac{1}{2mc^2}\left[E_n{}^2 + 2E_n\frac{e^2}{4\pi\varepsilon_0}\frac{1}{a_B n^2} + \left(\frac{e^2}{4\pi\varepsilon_0}\right)^2\frac{1}{a_B{}^2 n^3 l\left(l+\frac{1}{2}\right)}\right] =$$

$$= -E_n\frac{\alpha^2}{n^2}\left[\frac{3}{4} - \frac{n}{l+\frac{1}{2}}\right] \qquad (\boldsymbol{8.7.c})$$

where the $\alpha = e^2/4\pi\varepsilon_0 c\hbar$ is fine structure constant.

The calculation of spin-orbit interaction energy gives:

$$E_{SO} = -\frac{E_n}{2n}\alpha^2\frac{J(J+1) - l(l+1) - \frac{3}{4}}{l\left(l+\frac{1}{2}\right)(l+1)} \qquad (\boldsymbol{14.4.d})$$

where $J$ is the spin-orbit coupling quantum number given by $(l + s)$.

Finally, we have to calculate the Darwin energy:

$$E_{D_w} = \left\langle n,l,m\left|\frac{\pi\hbar^2}{2m^2c^2}\frac{e^2}{4\pi\varepsilon_0}\delta(r)\right|n,l,m\right\rangle = \begin{cases} 0 \;\; if \;\; l \neq 0 \\ -E_n\dfrac{\alpha^2}{n} \;\; if \;\; l = 0 \end{cases} \qquad (\boldsymbol{14.4.e})$$

The total energy (approximated) of the pure state $|n,l,m\rangle$ according the relativistic theory is:

$$E_{rel} = E_n - E_n \frac{\alpha^2}{n^2} \left[ \frac{3}{4} - \frac{n}{l + \frac{1}{2}} \right] - \frac{E_n}{2n} \alpha^2 \frac{J(J+1) - l(l+1) - \frac{3}{4}}{l \left( l + \frac{1}{2} \right) (l+1)} - E_n \frac{\alpha^2}{n} =$$

$$= \frac{E_n}{2n} \alpha^2 \left[ \frac{1}{n} \left( \frac{3}{4} - \frac{n}{l + \frac{1}{2}} \right) - \frac{1}{2} \frac{J(J+1) - l(l+1) - \frac{3}{4}}{l \left( l + \frac{1}{2} \right) (l+1)} - 1 \right] \qquad (\mathbf{14.4.f})$$

If $l \neq 0$ then the number $-1$ in the square bracket disappears. The relativistic approach proves that the energies of the states of the hydrogen atom depend not only by the main quantum number $n$, as instead occur in the non-relativistic quantum theory, but also by the numbers $l$ and $s$. As will be discussed in the next section, the **(14.4.f)** is of fundamental importance for the interpretation of the fine structure of hydrogen atom spectra.

It must be noted that the Thomas energy is always negative just like $E_n$ (all the electron states are tied). Therefore, the relativistic correction of the kinetic energy stabilizes the electronic states calculated by the Schrödinger equation. The spin-orbit interaction is not allowed for $l = 0$ (orbitals with spherical symmetry); that means the correction of the non-relativistic energy occurs only for the states with a non-spherical symmetry. Concerning the same term we are considering, we note that if $s = 1/2$ then $E_{SO}$ is positive while if $s = -1/2$ then it becomes negative. The Darwin energy is zero if $l \neq 0$ and is positive if $l = 0$; that means it gives always a destabilizing contribution to the states with spherical symmetry. Finally, we note that for $l = 0$ in the **(14.4.d)** appears a non-determined form $\frac{0}{0}$. However, the infinitesimal terms to the numerator and denominator have the same order so that:

$$\lim_{l \to 0} - \frac{E_n}{2n} \alpha^2 \frac{J(J+1) - l(l+1) - \frac{3}{4}}{l \left( l + \frac{1}{2} \right) (l+1)} = - \frac{E_n}{2n} \alpha^2$$

which is just the Darwin contribution to the total energy. The physical meaning of this result is the following: the positive part of the spin-orbit correction for the s-type orbitals is due to the Darwin correction.

## 14.5. The Fine Structure of the Hydrogen Atom

As said in the introductory section, the fine resolution of the hydrogen atom spectrum shows that the lines are formed by others thinner. This structure of the spectrum is defined *fine structure* and can be explained only by the use of relativistic quantum theory. The introduction of the spin-orbit interaction leads to explain the splitting of spectral lines; however, is thanks to the Dirac equation that the interpretation of the fine structure assumes its final form.

Let's consider the first level of the hydrogen atom represented by the state $|1, 0, 0\rangle$; its energy calculated by the Schroedinger equation is $E = -13.6\ eV$. According with the (**14.4.f**) the relativistic correction is:

$$\frac{1}{4}\alpha^2\left[\frac{1}{2}\left(\frac{3}{4} - 2\right) - 1\right]$$

that is obtained replacing $n = 1$ and omitting the spin-orbit interaction since $l = 0$. Multiplying the Scroedinger energy times this factor the state $|1, 0, 0\rangle$ is stabilized of $181 \cdot 10^{-6}\ eV$ (a very small quantity being the 0.0013 % of the uncorrect value, confirming the correctness of the perturbation theory approach we adopted in the previous section).

The second level of hydrogen atom is formed by four degenerate states $|2, 0, 0\rangle$, $|2, 1, 0\rangle$, $|2, 1, 1\rangle$ and $|2, 1, -1\rangle$; the Schrödinger energy of these orbitals is $E = -3.4\ eV$. Applying the relativistic theory a first splitting between the levels $2s$ and $2p$ occurs, due to the Thomas correction that stabilizes of $147.088 \cdot 10^{-6}\ eV$ the first one and of $26.4 \cdot 10^{-6}\ eV$ the second. The Darwin term, on the contrary, destabilizes only the orbital $2s$ of $90.5159 \cdot 10^{-6}\ eV$ since for the levels $2p$ the quantum number $l$ is 1. As a whole, the relativistic effects of Thomas and Darwin stabilize the orbital $2s$ of $56.5721 \cdot 10^{-6}\ eV$. For levels $2p$, being $l = 1$, is present also the spin-orbit interaction which leads to their splitting. In the case of orbital $2p$ with quantum number $J = 1/2$, it is stabilized of $30.172 \cdot 10^6\ eV$, while the orbitals $2p$ with $J = 3/2$ is destabilized of $15.086 \cdot 10^{-6}\ eV$. As a whole, the level $2p_{1/2}$ is stabilized of $56.572 \cdot 10^{-6}\ eV$, equal to the energy of the orbital $2s_{1/2}$, while the level $2p_{3/2}$ is stabilized of $11.28 \cdot 10^{-6}\ eV$.

For the third energy level we proceed following the same scheme adopted for $n = 2$, taking in consideration also the orbitals $3d$ that, being characterized by the quantum number $l = 2 \neq 0$, presents the spin-orbit interaction just like the orbitals $3p$. As a whole, the levels $3s$ and $3p$ with $J = 1/2$ are stabilized of $20.08 \cdot 10^{-6}\ eV$, the levels $3p$ and $3d$ with $J = 3/2$ are stabilized of $6.69 \cdot 10^{-6}\ eV$ and, finally, the orbitals with $J = 5/2$ are stabilized of $2.23 \cdot 10^{-6}\ eV$. In table are instead listed all the energy contributions of fine structure for the hydrogen atom yielded by the Dirac equation:

| Level | Schrödinger Energy (eV) | Spin-Orbit Energy $(10^{-6}\,eV)$ | Thomas Energy $(10^{-6}\,eV)$ | Darwin Energy $(10^{-6}\,eV)$ | $\Delta E_{rel-Sch}$ $(10^{-6}\,eV)$ |
|---|---|---|---|---|---|
| 1s1/2 | -13.598 | 0 | -905.159 | 724.128 | -181.032 |
| 2p1/2 | -3.399 | -30.172 | -26.400 | 0 | -56.572 |
| 2p3/2 | -3.399 | 15.086 | -26.400 | 0 | -11.314 |
| 2s1/2 | -3.399 | 0 | -147.088 | 90.515 | -56.572 |
| 3p1/2 | -1.510 | -8.939 | -11.174 | 0 | -20.114 |
| 3p3/2 | -1.510 | 4.469 | -11.174 | 0 | -6.704 |
| 3s1/2 | -1.510 | 0 | -46.934 | 26.819 | -20.114 |
| 3d3/2 | -1.510 | -2.681 | -4.022 | 0 | -6.704 |
| 3d5/2 | -1.510 | 1.787 | -4.022 | 0 | -2.234 |
| 4p1/2 | -0.849 | -3.771 | -5.421 | 0 | -9.193 |
| 4p3/2 | -0.849 | 1.885 | -5.421 | 0 | -3.535 |
| 4s1/2 | -0.849 | 0 | -20.507 | 11.314 | -9.193 |
| 4d3/2 | -0.849 | -1.131 | -2.404 | 0 | -3.535 |
| 4d5/2 | -0.849 | 0.754 | -2.404 | 0 | -1.650 |
| 4f5/2 | -0.849 | -0.538 | -1.111 | 0 | -1.650 |
| 4f7/2 | -0.849 | 0.404 | -1.111 | 0 | -0.707 |
| 5p1/2 | -0.543 | -1.931 | -2.993 | 0 | -4.924 |
| 5p3/2 | -0.543 | 0.965 | -2.993 | 0 | -2.027 |
| 5s1/2 | -0.543 | 0 | -10.717 | 5.793 | -4.924 |
| 5d3/2 | -0.543 | -0.579 | -1.448 | 0 | -2.027 |
| 5d5/2 | -0.543 | 0.386 | -1.448 | 0 | -1.062 |

**Table 7**

From the values listed we may state that:

- the sum of all relativistic corrections stabilize the Schrödinger levels with a trend that decreases with the main quantum number $n$. This behavior has expected because increasing the main quantum number the electron velocity decreases (see the expression of the velocity obtained applying the Bohr model);

- all relativistic corrections are smaller of several orders of magnitude than the Schrödinger energy (good conditions for the application of the perturbation theory);

- the Thomas energy is always negative (stabilizing effect);

- the Darwin energy is always positive for all s-type orbital and is zero when the quantum number $l = 0$.

Using the spectroscopic selection rules is possible to obtain all the allowed electron transitions between the states so to reproduce the whole structure of the spectrum. In figure 3 (Grotrian diagram) are represented the electron transitions giving the series of Lyman, Balmer and Paschen for the hydrogen atom:

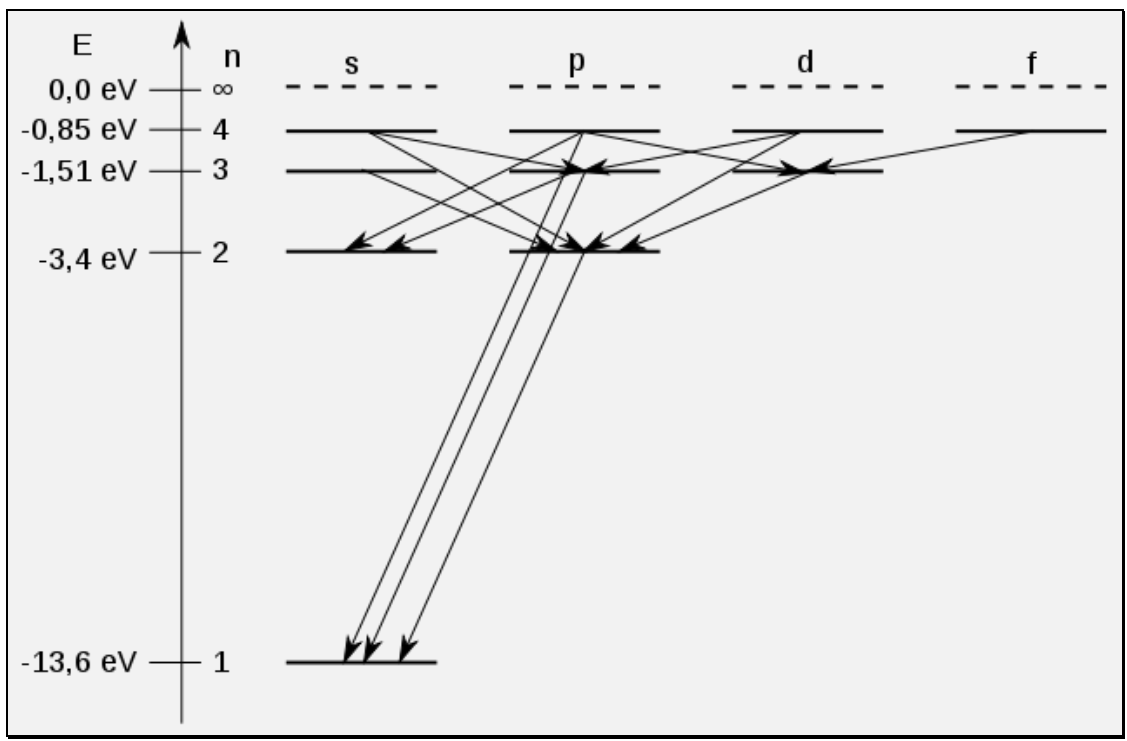

**Figure 3**

We note that none *perpendicular* transitions is present and only the transversal ones are allowed. This is in agreement with the spectroscopic rule $\Delta l = \pm 1$ which states that only transitions between orbitals with different geometry take place. The Lyman series is in the UV band of the spectrum and is produced by the transitions from the levels $np$ to the fundamental state $1s$; they are the most energetic. The Balmer series is in the VIS band and is due to the transitions from the orbitals $np$ to the $2s$, from the $ns$ to the $2p$ and from the $nd$ to the $2p$ (with $n > 2$). Finally, the Paschen series is in the IR band and is generated by the transitions from the orbitals $ns$, $np$, $nd$, with $n > 4$, to the orbitals $3s$, $3p$, $3d$.

## 15. Conclusion

The quantum theory has been developed with the aim of solving the problem of the spectrum of the hydrogen atom. The obtained equations, however, are solvable only for a limited number of real cases, like the hydrogen atom, the like-hydrogen ions and the free electron. In more complex cases is necessary to introduce approximations based on assumptions physically acceptable. This limitation, however, has enabled to develop a new branch of the atomic theory which deals to search new computational methods able to explain correctly the experimental results. Is not a case that quantum theory applied to the physical-chemistry has been able to explain with high precision the properties of atoms belonging to the Mendeleev table and of a large part of the molecules. This knowledge leads to design new compounds whose properties are studied for the development of new technologies. In the world we are living we use in every moment objects whose making was made possible thanks to the correctness of the equations that have been taken under review in this work. The quantum physics accounted for humanity the knowledge that allowed within a century a technological and economic development that classic physics has produced over a millennium.

### 16. References

[1] H. Cavendish – On Airs Fittizi (1766).

[2] A.J. Angstrom, "Recherches sur le Spectre Solaire, Spectre Normal du Soleil", Atlas de Six Planches, Upsal, W. Schultz, Imprimeur de l'universitè; Berlin, Ferdinand.

[3] J.J. Thompson, Philosophical Magazine 44, 295 (1897).

[4] J.J. Balmer, Annalen der Physik und Chemie 25, 80-85 (1885).

[5] F. Paschen, Annalen der Physik 332 (13) – 537-570 (1908).

[6] F. Brackett, Astrophysical Jurnal 56 – 154 (1922).

[7] A.H. Pfund, Jurnal Opt. Soc. Am. 9 (3) – 193-196 (1924).

[8] C.J. Humphreys, J. Research Nat. Bur. Standards 50 (1953).

[9] N. Bohr, Philosophical Magazine, Series 6, Vol. 26, 1-25 (July 1913).

[10] P. Weinberg, Phil. Mag. Letters, Vol. 86, n° 7, July 2006, 405-410.

[11] M. Beller, Quantum Dialogue: the Making of a Revolution, Chicago University Pres (1999).

[12] A. Sommerfeld, Annalen der Physik, 1916 [4] 51, 1-94.

[13] L. de Broglie, Ann. De Phys. 10e serie, t. III (Janvier-Fevrier 1925).

[14] L. de Broglie, J. De Physique (November 1922).

[15] E. Schrodinger, Annalen der Physik 79 (4), 734-756 (1926).

[16] E. Schrodinger, Annalen der Physik 79 (6), 489-527 (1926).

[17] E. Schrodinger, Annalen der Physik 80 (13), 437-490 (1926).

[18] E. Schrodinger, Annalen der Physik 81 (18), 109-139 (1926).

[19] E. Schrodinger, Physical Review 28 (6), 1049-1070 (1926).

[20] E. Schrodinger, Annalen der Physik 82 (2), 265-272 (1927).

[21] J. von Neumann, Mathematical Foundation of Quantum Mechanics – Princenton Paperbacks.

[22] W. Heisenberg, Zeitschr. F. Phys. (17), 1-26 (1927).

[23] M. Born, Gott. Naschr 1926, p. 146: AA Vol. 2, p. 284.

[24] M Born, Zeitschr. F. Phys. (37), 863, 1926; AA Vol. 2, p. 228.

[25] M Born, Zeitschr. F. Phys. (38), 803, 1926; AA Vol. 2, p. 233.

[26] W. Heisenberg, Zeitschr. F. Phys. (33), 879-893 (1925).

[27] M. Razavy, Heisenberg's Quantum Mechanics, World Scientific Publishing (2011).

[28] P.A.M. Dirac, The Quantum Theory of Electron, Proceeding of the Royal Society A: Mathematical, Physical and Engeneer Sciences 117 (778): 610 (1928).

[29] L. Landau, E. Lifszit, P. Pitaevskij, Relativistic Quantum Theory – Volume IV of the Theoretical Physics Course – Mir Edition 1975, Moscow.

[30] C. Anderson, Physical Review, Vol. 43, 491-498 (1033).

[31] W. Pauli, Collected Scientific Papers, Vol. 1,2 – Interscience Publisher (1964).